\documentclass[aps,twocolumn,nofootinbib,showpacs,prd,aps,10pt,superscriptaddress]{revtex4-2}
\usepackage[dvips]{graphicx}
\usepackage[english]{babel}
\usepackage[T1]{fontenc}
\usepackage{mathrsfs}
\usepackage[tbtags]{amsmath}
\usepackage{amssymb}
\usepackage{amsxtra}
\usepackage{amsopn}
\usepackage{latexsym}
\usepackage[mathcal]{eucal}
\usepackage{mathtools}
\usepackage{slashed}

\newcommand{\BE}{\begin{equation}}
\newcommand{\EE}{\end{equation}}
\newcommand{\BA}{\begin{align}}
\newcommand{\EA}{\end{align}}

\newcommand{\nn}{\nonumber}

\newcommand{\tx}{\text}
\newcommand{\mc}{\mathcal}
\newcommand{\psibar}{\overline{\psi}}
\newcommand{\cbar}{\overline{c}}
\newcommand{\avg}[1]{\left\langle#1\right\rangle}

\begin{document}

\title{Screened massive expansion of the quark propagator in the Landau gauge}

\author{Giorgio Comitini}
\email{giorgio.comitini@dfa.unict.it}
\affiliation{Dipartimento di Fisica e Astronomia ``E. Majorana'', Universit\`a di Catania, Via S. Sofia 64, I-95123 Catania, Italy}
\affiliation{INFN Sezione di Catania, Via S. Sofia 64, I-95123 Catania, Italy}

\author{Daniele Rizzo}
\email{daniele.rizzo@studium.unict.it}
\affiliation{Dipartimento di Fisica e Astronomia ``E. Majorana'', Universit\`a di Catania, Via S. Sofia 64, I-95123 Catania, Italy}

\author{Massimiliano Battello}
\email{massimiliano.battello@gmail.com}
\affiliation{Dipartimento di Fisica e Astronomia ``E. Majorana'', Universit\`a di Catania, Via S. Sofia 64, I-95123 Catania, Italy}

\author{Fabio Siringo}
\email{fabio.siringo@ct.infn.it}
\affiliation{Dipartimento di Fisica e Astronomia ``E. Majorana'', Universit\`a di Catania, Via S. Sofia 64, I-95123 Catania, Italy}
\affiliation{INFN Sezione di Catania, Via S. Sofia 64, I-95123 Catania, Italy}

\date{\today}

\begin{abstract}
The infrared behavior of the quark propagator is studied at one loop and in the Landau gauge ($\xi=0$) using the screened massive expansion of full QCD and three different resummation schemes for the quark self-energy. The shift of the expansion point of perturbation theory, which defines the screened expansion, together with a non-standard renormalization of the bare parameters, proves sufficient to describe the dynamical generation of an infrared quark mass also in the chiral limit. Analytically, the scale for such a mass is set by a mass parameter $M$, whose value is fixed by a fit to the lattice data for quenched QCD. The quark mass function $\mathcal{M}(p^{2})$ is shown to be in very good agreement with the lattice results. The quark $Z$-function, on the other hand, shows the wrong qualitative behavior in all but one of the studied resummation schemes, where its behavior is qualitatively correct, but only at sufficiently high energies.
\end{abstract}



\maketitle
\section{Introduction}

In the Standard Model of particle physics the light quarks acquire their masses dynamically through two separate and complementary mechanisms. The first one is the spontaneous breaking of the electroweak gauge symmetry $U(1)_{Y}\times SU(2)_{L}$, induced by a non vanishing vacuum expectation value (VEV) for the Higgs field. Due to the former, a quark mass $M_{q}$ is generated which is proportional to the product of the quark-Higgs Yukawa coupling and the Higgs field VEV. The second mechanism is a remnant of the violation of global chiral symmetry. In this context, the violation is caused by the strong interactions and manifests itself in a non-zero VEV for the quark mass operator $\psibar\psi$ -- i.e. of the quark condensate --, which would be constrained to vanish in the presence of chiral symmetry. In turn, the quark condensate triggers the non-vanishing of the quark mass function $\mc{M}(p^{2})$ in the chiral limit, as can be proven by an operator product expansion (OPE) of the quark propagator. Despite being obeyed by the massless quarks only, limited to the light quarks ($M_{q}\ll \Lambda_{\tx{QCD}}$, where $\Lambda_{\tx{QCD}}$ is the QCD scale) chiral symmetry is still a good approximate symmetry of the QCD Lagrangian; the mechanism that underlies its violation leads to the dressing of the light Higgs-generated masses, greatly enhancing their effective values in the infrared (IR) regime.

Studying the origin of the quark effective masses in the IR is of paramount importance for understanding the experimentally observed hadron spectrum. This is rooted in the fact that the measured values of the light Higgs-generated masses -- $M_{u}\approx 2.2$~MeV, $M_{d}\approx 4.7$~MeV, $M_{s}\approx 93$~MeV for the up, down and strange quarks respectively \cite{rpp} -- do not compare well with the observed values of the (unflavored) baryon masses, which are of the order of $1$~GeV. The infrared enhancement, induced by the violation of chiral symmetry, is a good candidate for filling the gap between those masses. Unfortunately, mainly because of the non-perturbative nature of dynamical mass generation, no purely analytical and fully predictive description of the latter in the framework of first principles QCD is available to date.

In the context of the strong interactions, dynamical mass generation has been an active field of research for decades now. The development of chiral perturbation theory in the 1960s and 1970s offered a framework in which the large observed masses of the hadrons could be understood to be a consequence of chiral symmetry violation. In the gauge sector, the hypothesis that the gluons might acquire an infrared mass as a result of their self-interactions was advanced by Cornwall in 1982 \cite{cornwall} and confirmed by lattice studies in the 2000s \cite{stern07,olive,cucch07,cucch08,cucch08b,cucch09,bogolubsky,olive09,dudal,binosi12,olive12,burgio15,duarte}. In the continuum, considerable progresses have been made by the numerical integration of integral equations \cite{aguilar8,aguilar10,aguilar14,aguilar14b,papa15,papa15b,huber14,huber15g,huber15b,huber20,mitter,cyrol16,cyrol18}, by variational methods \cite{reinhardt04,reinhardt05,reinhardt08,reinhardt14,sigma,sigma2,gep2,varqed,varqcd,genself,ptqcd0}, and by physically motivated phenomenological models \cite{GZ,shakinPRD,iparticle,tissier10,tissier11,tissier14,sorella15,dudal15,dudal08,dudal08b,dudal11,machado}. For a recent review on the subject see Ref.~\cite{huberrev}. The generation of a mass for the gluons is of special interest from a theoretical point of view, since gauge invariance in the framework of ordinary perturbation theory (PT) forbids the gluons to acquire a mass.

While in principle the failure of ordinary PT to describe the gluon's infrared mass could be attributed to its break down at low energies, in recent years a new approach to the perturbation theory of pure Yang-Mills (YM) theory has shown that most of the non-perturbative content of the gluon dynamic -- at least as far as the two-point functions are concerned -- can be absorbed into a shift of the expansion point of the Yang-Mills perturbative series. This approach, termed the screened massive expansion \cite{ptqcd,ptqcd2,damp,varT,thermal,analyt,scaling,xigauge,xighost,beta,rg,tesim}, is a simple extension of ordinary PT, formulated in such a way as to treat the transverse gluons as massive already at tree level while leaving the total action of the theory unchanged. The screened expansion has proven to be self-consistent to one loop -- since it is renormalizable and leads to an infrared-finite and moderately small running coupling constant \cite{rg} -- and predictive when optimized by principles of gauge invariance \cite{xigauge}; it yields two-point functions which are in excellent agreement with the lattice data in the Landau gauge \cite{xigauge,rg}.

The main objective of this paper is to extend the formalism of the screened massive expansion to full QCD with one flavor of quark, with the aim of studying the infrared behavior of the quark propagator. The method was already applied in Refs.~\cite{analyt,scaling} to describe some of the low-energy features of the quark dynamics in the chiral limit; here we refine its definition, implement some of our latest findings on the gauge sector, extend the study to non-chiral quarks, and use a new set of lattice data as a benchmark for comparison and in order to fix some of the free parameters in our expressions.

Our treatment of the quark sector will closely follow what we did in pure Yang-Mills theory for the gluons; namely, we will shift the expansion point of the perturbative series by introducing a new mass parameter $M$ for the zero-order quark propagator. The motivation for the shift lies in the phenomenon of dynamical mass generation for the light quarks: as previously discussed, due to the strong interactions, at low energies the light quarks propagate with a mass which is greatly enhanced with respect to their tree-level (Lagrangian) value; since this effect cannot be captured by ordinary perturbation theory, some kind of non-ordinary and non-perturbative resummation of the quark self-energy is needed in order to successfully describe the infrared quark dynamics. This is precisely what the shift does: by replacing the mass contained in the standard zero-order propagator with an enhanced mass parameter, it optimizes the expansion point of perturbation theory so that the quarks propagate with an effective infrared mass of the order of the QCD scale $\Lambda_{\tx{QCD}}$, rather than with the mass contained in the Lagrangian -- which would be more relevant to the high energy regime. The same is done for the transverse gluons, which at tree level are set up to propagate with a finite non-zero mass.

The shift is performed in such a way as to leave the total action of the theory unchanged. As a result, three new two-point interaction vertices arise which are proportional to the quark mass parameter $M$ and bare mass $M_{B}$ and to the gluon mass parameter $m^{2}$. Since the expansion cannot be carried out exclusively in powers of the coupling constant, the approach is non-perturbative in nature; nonetheless, the calculations are done using standard Feynman diagram techniques, so that the method is still perturbative in the widest sense of the word.

As we will see in the following sections, our analysis still has major theoretical limitations. First and foremost, the value of the quark mass parameter $M$ introduced by the shift needs to be fixed from external inputs in order to obtain definite quantitative results. At variance with pure Yang-Mills theory, where the method was optimized based on principles of gauge invariance and the redundancy in the number of free parameters was effectively eliminated (see Ref.~\cite{xigauge} and the discussion in Sec.~II), at this moment no such procedure is available for full QCD. Because of this, in order to test the strength of the screened expansion of QCD, we will resort to fitting the free parameters of the expansion using the lattice data; for reasons which will be discussed in a later section, the fit will be done using a set of data for quenched QCD.

Our study of the quark propagator will make use of three different resummation schemes for the quark self-energy: the minimalistic, vertex-wise and complex-conjugate schemes (to be defined in Sec.~III). The first and second ones are a variation on the same theme and only differ by the number of gluon mass counterterms (i.e. two-point mass vertices, see the next section) included in the computation of the self-energy. The complex-conjugate scheme, on the other hand, uses the fully dressed gluon propagator (or, to be precise, an approximation thereof) in place of the zero-order gluon propagator as the internal gluon line of the self-energy. Each of these schemes has strengths and weaknesses which will be discussed. For the moment, we anticipate that the three resulting mass functions $\mc{M}(p^{2})$ do not show significant differences and are in very good agreement with the lattice data (provided of course that the values of the free parameters are chosen appropriately). The quark $Z$-functions, conversely, show the wrong qualitative behavior in all but the complex-conjugate scheme; when computed using the latter, $Z(p^{2})$ is qualitatively correct at sufficiently high energies, but fails nonetheless at low energies.

Ultimately, we were not able to quantitatively reproduce the lattice $Z$-function using the method presented in this study. However, it must be kept in mind that, in the Landau gauge, the divergent part of the $Z$-function is exactly zero at one loop, and above $1.0-1.5$~GeV the finite contribution to $Z(p^{2})-1$ is quite small, yielding an almost constant $Z(p^{2})\approx 1$. Thus, the $Z$-function seems to be very sensitive to corrections coming from higher loops \cite{barrios21}, thermal effects \cite{olive19}, neglected non-perturbative terms and -- on the lattice side -- even artifacts which may affect the actual result found in the numerical simulations.\\

This paper is organized as follows. In Sec.~II we review the setup and results of the screened expansion of pure Yang-Mills theory; in Sec.~III we formalize the screened expansion of full QCD with one flavor of quark, discuss its renormalization and define the resummation schemes which we will use for the computation of the one-loop quark self-energy; in Sec.~IV we present our results for the quark propagator, fitting the free parameters of the expansion from the lattice data; in Sec.~V we discuss our results and present our conclusions.
\newpage

\section{The screened massive expansion of pure Yang-Mills theory}

The screened massive expansion for the gauge-fixed, renormalized Faddeev-Popov Lagrangian was developed in Refs.~\cite{ptqcd,ptqcd2} and extended to finite temperature in~\cite{damp,varT,thermal}, to full QCD in~\cite{analyt,scaling} and to a generic covariant gauge in~\cite{xigauge,xighost}. Its renormalization in the Landau gauge was discussed in Refs.~\cite{beta,rg}, where different renormalization schemes were considered and analytical expressions were reported for its beta function. The method has proven to be self-consistent and predictive when optimized by principles of gauge invariance \cite{xigauge,rg}.

In what follows we give a brief review of the setup and main results of the screened expansion of pure Yang-Mills theory in the Landau gauge. Both of these are functional to our analysis of full QCD.

\subsection{Setup of the method}

The bare Faddeev-Popov (FP) Lagrangian for pure SU(N) Yang-Mills theory in a general covariant gauge is given by
\BE\label{fplag}
\mc{L}=\mc{L}_{\tx{YM},B}+\mc{L}_{\tx{fix},B}+\mc{L}_{\tx{FP},B},
\EE
where
\begin{align}\label{barelagrangianterms}
\mc{L}_{\tx{YM},B}=-\frac{1}{2}\,\text{Tr}\left(F_{B\mu\nu}F^{\mu\nu}_{B}\right),\nn\\
\mc{L}_{\tx{fix},B}=-\frac{1}{\xi_{B}}\tx{Tr}\left(\partial_{\mu}A_{B}^{\mu}\partial_{\nu}A_{B}^{\nu}\right),\nn\\
\mc{L}_{\tx{FP},B}=\partial_{\mu}\cbar^{a}_{B}D^{\mu}_{B}c^{a}_{B}.
\end{align}
Here we have defined the bare gauge field $A^{\mu}_{B}$ as
\BE
A^{\mu}_{B}=A^{a\mu}_{B}T_{a},
\EE
where the $T_{a}$'s are SU(N) generators, chosen so that
\BE
\tx{Tr}\left(T_{a}T_{b}\right)=\frac{1}{2}\,\delta_{ab};
\EE
$\xi_{B}$ is the bare gauge parameter defining the covariant gauge and $F^{\mu\nu}_{B}$ is the bare field-strength tensor,
\BE
F_{B}^{a\mu\nu}=\partial^{\mu}A^{a\nu}_{B}-\partial^{\nu}A^{a\mu}_{B}+g_{B}f^{a}_{bc}A^{b\mu}_{B}A^{c\nu}_{B},
\EE
with
\BE
[T_{a},T_{b}]=if_{ab}^{c}T_{c}.
\EE
The bare covariant derivative $D_{B}^{\mu}$ acting on the ghost and antighost fields $c_{B}^{a},\overline{c}_{B}^{a}$ reads
\BE
(D^{\mu}_{B})^{a}_{c}=\delta^{a}_{c}\partial^{\mu}+g_{B}f^{a}_{bc}A^{b\mu}_{B}.
\EE
$\mathcal{L}$ can be renormalized by introducing suitable renormalization factors $Z_{A}$, $Z_{c}$ and $Z_{A\cbar c}$ for the gauge and ghost fields and for the coupling constant, respectively, and by defining new, renormalized gauge and ghost fields $A_{\mu}^{a}$, $c^{a}$ and $\cbar^{a}$, a renormalized coupling $g$ and a renormalized gauge parameter $\xi$, according to
\begin{align}
\nn A^{\mu}_{B}=Z_{A}^{1/2}A^{\mu},\quad \xi_{B}=Z_{A}\xi,\\
\nn c^{a}_{B}=Z_{c}^{1/2}c^{a},\quad \cbar^{a}_{B}=Z_{c}^{1/2}\cbar^{a},\\
g^{2}=g_{B}^{2}\,\frac{Z_{A}Z_{c}^{2}}{Z_{A\cbar c}^{2}}.
\end{align}
In terms of the renormalized fields, the Faddeev-Popov Lagrangian reads
\BE
\mc{L}=\mc{L}_{\tx{YM}}+\mc{L}_{\tx{fix}}+\mc{L}_{\tx{FP}}+\mc{L}_{\tx{c.t.}},
\EE
where
\begin{align}
\mc{L}_{\tx{YM}}=-\frac{1}{2}\,\text{Tr}\left(F_{\mu\nu}F^{\mu\nu}\right),\nn\\
\mc{L}_{\tx{fix}}=-\frac{1}{\xi}\tx{Tr}\left(\partial^{\mu}A_{\mu}\partial^{\nu}A_{\nu}\right),\nn\\
\mc{L}_{\tx{FP}}=\partial^{\mu}\cbar^{a}D_{\mu}c^{a},
\end{align}
and $\mc{L}_{\tx{c.t.}}$ contains the renormalization counterterms. The renormalized field-strength tensor $F_{\mu\nu}^{a}$ and covariant derivative $D_{\mu}$ are defined as
\begin{align}
\nn F^{a}_{\mu\nu}=\partial_{\mu}A^{a}_{\nu}-\partial_{\nu}A^{a}_{\mu}+gf^{a}_{bc}A^{b}_{\mu}A^{c}_{\nu},\\
(D_{\mu})^{a}_{c}=\delta^{a}_{c}\partial_{\mu}+gf^{a}_{bc}A^{b}_{\mu}.
\end{align}
We note that $\mc{L}_{\tx{c.t.}}$ does not contain a counterterm for the gauge-fixing term $\mc{L}_{\tx{fix}}$; indeed, the Slavnov-Taylor identities ensure that the bare gauge parameter $\xi_{B}$ can be multiplicatively renormalized by the gauge field renormalization factor $Z_{A}$ alone.

Ordinary perturbation theory is defined by a split of the renormalized Lagrangian,
\BE
\mc{L}=\mc{L}_{0}+\mc{L}_{\tx{int}}+\mc{L}_{\tx{c.t.}},
\EE
where $\mc{L}_{0}=\lim_{g\to 0}\mc{L}$ is taken to be the non-interacting limit of $\mc{L}$,
\BE
\mc{L}_{0}=\frac{1}{2}A_{\mu}^{a}\left[i\Delta^{\mu\nu}_{0ab}(p)^{-1}\right]A_{\nu}^{b}+\cbar^{a}\left[i\mc{G}_{0ab}(p^{2})^{-1}\right]c^{b};
\EE
here the ordinary zero-order gluon and ghost propagators $\Delta_{0\mu\nu}^{ab}$ and $\mc{G}^{ab}_{0}$ read
\begin{align}
\Delta_{0\mu\nu}^{ab}(p)=\frac{-i\delta^{ab}}{p^{2}}\,\left(t_{\mu\nu}(p)+\xi \ell_{\mu\nu}(p)\right),\nn\\
\mc{G}_{0}^{ab}(p^{2})^{-1}=\frac{i\delta^{ab}}{p^{2}},
\end{align}
where $t_{\mu\nu}(p)$ and $\ell_{\mu\nu}(p)$ are the transverse and longitudinal projectors defined as
\BE
t_{\mu\nu}(p)=\eta_{\mu\nu}-\frac{p_{\mu}p_{\mu}}{p^{2}},\quad \ell_{\mu\nu}(p)=\frac{p_{\mu}p_{\nu}}{p^{2}}.
\EE
The interaction term $\mc{L}_{\tx{int}}$ contains a three-gluon, a four-gluon and a ghost-gluon interaction,
\BE
\mc{L}_{\tx{int}}=\mc{L}_{3g}+\mc{L}_{4g}+\mc{L}_{\cbar cg},
\EE
where
\begin{align}\label{vertices}
\nn\mc{L}_{3g}&=-gf^{a}_{bc}\partial_{\mu}A_{\nu}^{a}A^{b\mu}A^{c\nu},\\
\nn\mc{L}_{4g}&=-\frac{1}{4}gf^{a}_{bc}f^{a}_{de}A_{\mu}^{b}A_{\nu}^{c}A^{d\mu}A^{e\nu},\\
\mc{L}_{\cbar cg}&=gf^{a}_{bc}\partial^{\mu}\cbar^{a}A_{\mu}^{b}c^{c}.
\end{align}
On the other hand, the term $\mc{L}_{\tx{c.t.}}$ contains the field and coupling renormalization counterterms,
\BE
\mc{L}_{\tx{c.t.}}=-\frac{1}{2}\delta_{A}\delta_{ab}p^{2}t^{\mu\nu}(p)A_{\mu}^{a}A^{b}_{\nu}+\delta_{c}\delta_{ab}p^{2}\cbar^{a}c^{b}+\cdots,
\EE
where $\delta_{A}=Z_{A}-1$ and $\delta_{c}=Z_{c}-1$. In particular, the gluon field renormalization counterterm is completely transverse.

At low energies, the ordinary perturbation theory of pure YM theory is known to be inconsistent due to the presence of an IR Landau pole in the running of the strong coupling constant. Moreover, constraints due to gauge invariance -- when applied in the framework of ordinary perturbation theory -- prevent the generation of an IR dynamical mass for the gluons, a phenomenon which by now has been well established mainly thanks to lattice calculations \cite{olive,cucch07,cucch08,cucch08b,cucch09,bogolubsky,olive09,dudal,binosi12,olive12,burgio15,duarte}. Addressing these issues is the main objective of the screened massive expansion.\\

The screened massive expansion of pure YM theory is defined by a shift of the expansion point of the Yang-Mills perturbative series, performed in such a way as to treat the transverse gluons as massive already at tree-level \cite{ptqcd,ptqcd2}. Explicitly, a shifting term $\delta\mc{L}$ is added to the zero-order (kinetic) part of the gauge-fixed, renormalized Fadeev-Popov Lagrangian, and subtracted back from its interaction part,
\BE
\mc{L}_{0}^{\prime}=\mc{L}_{0}+\delta \mc{L},\quad\mc{L}_{\tx{int}}^{\prime}=\mc{L}_{\tx{int}}-\delta \mc{L};
\EE
$\delta\mc{L}$ is chosen so that the zero-order transverse gluon propagator contained in $\mc{L}_{0}^{\prime}$ is replaced by a massive one: in momentum space
\BE
\delta \mc{L}=\frac{1}{2}A_{\mu}^{a}(p)i\left[i\Delta^{-1\,\mu\nu}_{m\,ab}(p)-i\Delta^{-1\,\mu\nu}_{0\,ab}(p)\right]A_{\nu}^{b}(-p),
\EE
where
\BE\label{glupropzero}
\Delta_{m\,ab}^{\mu\nu}(p)=\delta_{ab}\left\{\frac{-it^{\mu\nu}(p)}{p^{2}-m^{2}}+\xi\frac{-i\ell^{\mu\nu}(p)}{p^{2}}\right\}
\EE
is the new, massive zero-order gluon propagator. Since $\delta\mc{L}$ is added to and subtracted from the FP Lagrangian, the shift does not not modify the full action of Yang-Mills theory. Instead, it introduces a new free mass parameter $m^{2}$ and changes the Feynman rules of YM theory in two respects. First of all, since the new zero-order Lagrangian $\mc{L}_{0}^{\prime}$ reads
\BE
\mc{L}_{0}^{\prime}=\frac{1}{2}A_{\mu}^{a}\left[i\Delta^{-1\,\mu\nu}_{m\,ab}\right]A_{\nu}^{b}+\cbar^{a}\left[i\mc{G}_{0ab}^{-1}\right]c^{b},
\EE
the transverse gluons propagate with a massive propagator rather than with a massless one -- see Eq.~\eqref{glupropzero}. Second of all, the interacting part of the Lagrangian, $\mc{L}_\tx{int}^{\prime}$, contains a new two-point interaction, namely
\BE
-\delta\mc{L}=-\frac{1}{2}A_{\mu}^{a}(p)\,i\delta\Gamma_{g\ ab}^{\mu\nu}(p)\,A_{\nu}^{b}(-p),
\EE
where the vertex $\delta\Gamma_{g\ ab}^{\mu\nu}(p)$ is given by
\BE\label{masscount}
\delta\Gamma^{\mu\nu}_{g\ ab}(p)=-im^{2}t^{\mu\nu}(p)\delta_{ab}.
\EE
We refer to the latter as the gluon mass counterterm, not to be confused with the renormalization counterterms contained in $\mc{L}_{\tx{c.t.}}$. Neither the remaining interaction vertices -- spelled out in Eq.~\eqref{vertices} -- nor the renormalization counterterms are affected by the shift.
\begin{figure}[t] \label{fig:diagrams}
\centering
\includegraphics[width=0.25\textwidth,angle=-90]{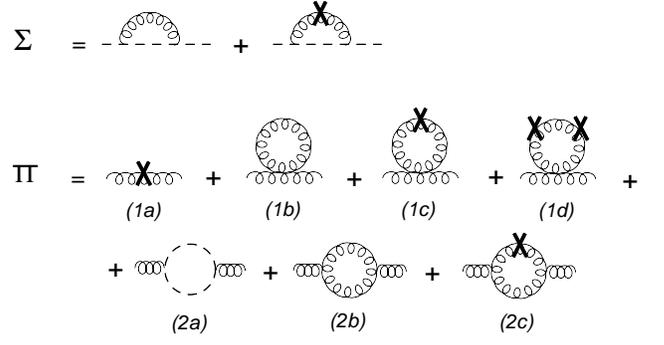}
\caption{Two-point graphs with no more than three vertices and no more than one loop. 
The cross is the transverse mass counterterm of Eq.~\eqref{masscount} and is regarded as
a two-point vertex. The renormalization counterterms are not shown in the figure.}
\end{figure}

The quantities of physical interest can be computed in the framework of the screened expansion using the Feynman rules described above. Since the vertex $\delta\Gamma_{g}$ is not proportional to the coupling constant, diagrams with an arbitrary number of vertices -- termed crossed diagrams if they contain at least one gluon mass counterterm -- coexist at any given loop order. For this reason, the screened expansion is intrinsically non-perturbative.

The crossed diagrams can be computed as derivatives of non-crossed diagrams with respect to the gluon mass parameter. This easily follows from the equality \cite{tesim}
\begin{align}\label{gluonders}
\nn\left[\Delta_{m}(p)\cdot \left(\delta\Gamma_{g}(p)\cdot\Delta_{m}(p)\right)^{n}\right]^{\mu\nu}_{ab}=\\
\nn=\frac{-i (-m^{2})^{n}}{(p^{2}-m^{2})^{n+1}}\,t^{\mu\nu}(p)\,\delta_{ab}=\\
=\frac{(-m^{2})^{n}}{n!}\,\frac{\partial^{n}}{\partial (m^{2})^{n}}\ \Delta_{m\,ab}^{\mu\nu}(p),
\end{align}
which is valid for every $n\geq 1$ and in any covariant gauge, and carries over to the loop integrals.

Due to the massiveness of the zero-order gluon propagator in the screened expansion, new UV divergences arise in the loop integrals which are proportional to the gluon mass parameter $m^{2}$. These divergences do not invalidate the renormalizability of the $n$-point functions of the theory, since they cancel as soon as crossed diagrams with a higher number of crossed vertices are taken into account \cite{ptqcd2,tesim}. The removal of mass divergences can (and indeed must) be adopted as a criterion for fixing the minimum number of crossed loops to be included when computing some quantity at a given loop order \cite{ptqcd2,tesim}.\\

To one loop, the one-particle-irreducible (1PI) gluon polarization $\Pi_{\mu\nu}^{ab}(p)$ and ghost self-energy $\Sigma^{ab}(p^{2})$ were computed from the diagrams in Fig.~1. The crossed vertices in the figure represent the gluon mass counterterm $\delta\Gamma_{g}$. Diagrams (1c) and (2c) in the gluon polarization are required in order to eliminate the mass divergences that arise from diagrams (1b) and (2b), respectively; they have a total of three vertices. To one loop, there are two more diagrams with the same number of vertices -- namely, diagram (1d) and the crossed diagram in the ghost self-energy (top right diagram in Fig.~1); these were also included in the one-loop calculation for consistency.

Since the shift that defines the screened expansion does not change the total action of pure YM theory, the full 1PI gluon polarization is known to be transverse by the Slavnov-Taylor identities. Therefore we can write
\BE
\Pi_{\mu\nu}^{ab}(p)=\Pi(p^{2})\,t_{\mu\nu}(p)\delta^{ab},
\EE
where $\Pi(p^{2})$ is the gluon's scalar polarization. After the resummation of the 1PI diagrams, the transverse-gluon and ghost dressed propagators $\Delta(p^{2})$ and $\mc{G}(p^{2})$ can then be expressed as
\begin{align}
\Delta(p^{2})=-i[p^{2}-m^{2}-\Pi(p^{2})]^{-1},\nn\\
\mc{G}(p^{2})=i[p^{2}-\Sigma(p^{2})]^{-1},
\end{align}
where $\Sigma(p^{2})$ is the ghost self-energy. Diagram (1a) in Fig.~1 is easily shown to contribute to the gluon polarization with a constant term $\Delta\Pi=-m^{2}$,
\BE
\Pi(p^{2})=-m^{2}+\Pi^{(\tx{loops})}(p^{2}),
\EE
where $\Pi^{(\tx{loops})}(p^{2})$ is the loop contribution to the polarization -- diagrams (1b) to (2c) in Fig.~1. It is then easy to see that the tree-level mass term inherited from the shift cancels out with $\Delta\Pi$, so that the dressed propagator itself can be expressed as
\BE
\Delta(p^{2})=-i[p^{2}-\Pi^{(\tx{loops})}(p^{2})]^{-1}.
\EE
From the above equation it is clear that in the screened expansion, rather than being a trivial effect of the shift of the expansion point, the gluon mass must come from the loops and is thus genuinely dynamical in nature; it does not coincide with the gluon mass parameter $m^{2}$, which at this stage is just an undetermined dimensionful scale.

Quite interestingly, the existence of a finite mass-scale in YM theory has been derived in the Gaussian approximation from first principles \cite{varT,tesim} but, of course, the actual value of that scale can only arise from the phenomenology, since there is no energy scale in pure YM theory. The best variational Gaussian vacuum
was shown to be the vacuum of a massive gluon, and the present screened expansion emerged has the perturbative loop expansion around that best massive vacuum \cite{varT}. While fermions have also been incorporated in the Gaussian formalism in the past \cite{HiggsTop}, it is not clear if the screened expansion of full QCD, as is discussed in the present paper, can also be regarded as a loop expansion around a variational Gaussian vacuum which breaks the chiral symmetry.

\subsection{Optimization and results in the Landau gauge}

In a general renormalization scheme and in the Landau gauge, the dressed gluon propagator $\Delta(p^{2})$ can be expressed as
\BE\label{gluprop}
\Delta(p^{2})=\frac{-iZ_{\Delta}}{p^{2}(F(s)+F_{0})},
\EE
where $s=-p^{2}/m^{2}$ and $Z_{\Delta}$ and $F_{0}$ are, respectively, a multiplicative and an additive renormalization constant\footnote{The strong coupling constant $\alpha_{s}$ was absorbed into the definition of $Z_{\Delta}$ and $F_{0}$, and makes no explicit appearance in what follows.}.  The function $F(s)$ was computed to one loop and third order in the number of vertices from the diagrams in Fig.~1; its analytical expression is reported in Ref.~\cite{ptqcd2}. The zero-momentum limit of $F(s)$ reads
\BE
F(s)\to\frac{5}{8s}\quad(s\to 0),
\EE
so that
\BE
\Delta(p^{2})\to\frac{i8Z_{\Delta}}{5m^{2}}\quad(p^{2}\to 0),
\EE
implying that the screened expansion's gluon propagator is indeed massive in the infrared. We reiterate that the gluon mass -- as defined, for instance and non-univocally, by $i\Delta(0)^{-1}$ -- comes from the loops and is thus dynamical in nature.

Together with the gluon mass parameter $m^{2}$, $Z_{\Delta}$ and $F_{0}$ are the only free parameters determining the gluon propagator in the screened expansion. The multiplicative constant $Z_{\Delta}$ can of course be fixed by renormalizing the propagator at some specified renormalization scale $p^{2}=-\mu^{2}$, i.e. by requiring that
\BE
\Delta(-\mu^{2})=\frac{-i}{-\mu^{2}}.
\EE
The value of the additive renormalization constant $F_{0}$, on the other hand, was optimized and fixed in Ref.~\cite{xigauge} according to principles of gauge invariance. In more detail, it was shown that there exists a value of $F_{0}$ in the Landau gauge, namely $F_{0}=-0.876$, which, when evolved to a general covariant gauge ($\xi\neq 0$), yields gauge-invariant poles $p_{0}^{2}$ for the gluon propagator whose residues are also gauge invariant in phase to less than $0.3\%$ \cite{nielsen,kobes90,breck}.

\begin{table}[t]
\setlength{\tabcolsep}{4pt}
\begin{tabular}{c||c|c||c}
$F_{0}$&$z_{0}^{2}$&$\varphi$&$p_{0}$ (GeV)\\
\hline
\hline
$-0.876$&$0.4575\pm 1.0130\,i$&$\pm\,1.262$&$\pm0.5810\pm0.3571\,i$
\end{tabular}
\caption{Results of the screened massive expansion of pure YM theory, obtained by imposing the gauge-parameter independence of the poles and of the phases of the residues of the gluon propagator in a general covariant gauge. From left to right: the additive renormalization constant $F_{0}$ in the Landau gauge; the adimensional position $z_{0}^{2}=p_{0}^{2}/m^{2}$ of the poles of the gluon propagator in the Landau gauge; the gauge-invariant phases $\varphi$ of the residues of the gluon propagator; the gauge-invariant dimensionful positions of the poles of the propagator, assuming $m=0.6557$~GeV in the Landau gauge (the $\pm$ signs are independent from each other).}
\end{table}

In the same context (and in previous papers also, see e.g. \cite{analyt,damp}), we found that the screened expansion's gluon propagator has two complex-conjugate poles, whose adimensional positions $z_{0}^{2}=p_{0}^{2}/m^{2}$ and $\overline{z_{0}^{2}}$ were determined in \cite{xigauge} from first principles. The existence of complex-conjugate poles has been related in the literature to the issue of the violation of positivity of the gluon spectral function and, more generally, to that of confinement \cite{stingl,kondo2021}. The poles and phases of the residues of the gluon propagator, as computed in the (optimized) screened expansion, are reported in Tab.~I.

Of particular relevance to this paper is the fact that the principal part of the gluon propagator -- i.e. the term which contains its poles -- well-approximates the full propagator itself~\cite{tesim}, provided that the former is multiplied by a factor of 0.945. This is shown in Fig.~2.

\begin{figure}[t]
\vskip 1cm
\centering
\includegraphics[width=0.30\textwidth,angle=-90]{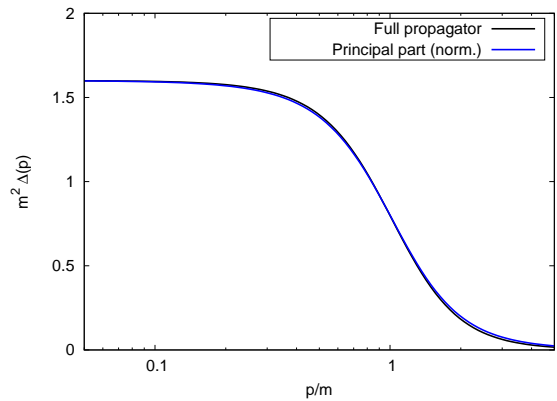}
\caption{Transverse gluon propagator in the Landau gauge ($\xi=0$) and in Euclidean space, computed in the screened expansion of pure YM theory. Black line: full one-loop propagator. Blue line: principal part of the one-loop propagator, normalized by a factor of 0.945.}
\end{figure}
\

With $Z_{\Delta}$ and $F_{0}$ fixed, the gluon mass parameter $m^{2}$ is left as the only free parameter of the theory (at least as far as the gluon two-point function is concerned). $m^{2}$ sets the energy units for the dimensionful quantities in the theory; as such, it cannot be determined from first principles and must be fixed from phenomenology. In this respect, the gluon mass parameter plays the same role as the QCD scale $\Lambda_{\tx{QCD}}$ of ordinary perturbation theory\footnote{For a lengthy discussion on the conceptual similarities between the gluon mass parameter $m^{2}$ and the QCD scale $\Lambda_{\tx{QCD}}$ see Ref.~\cite{rg}, where the issue was addressed in the context of the renormalization group improvement of the screened expansion.}. The propagator defined by Eq.~\eqref{gluprop}, with $F_{0}=-0.876$ optimized by principles of gauge invariance, was found to accurately reproduce the Euclidean lattice data of Ref.~\cite{duarte}, provided that the energy units of the screened expansion are set by choosing $m=0.6557$~GeV (see Fig.~3). Once the value of the gluon mass parameter is determined, the dimensionful values of the poles of the propagator can be computed; they are reported in the last column of Tab.~I.

\begin{figure}[t]
\vskip 1cm
\centering
\includegraphics[width=0.30\textwidth,angle=-90]{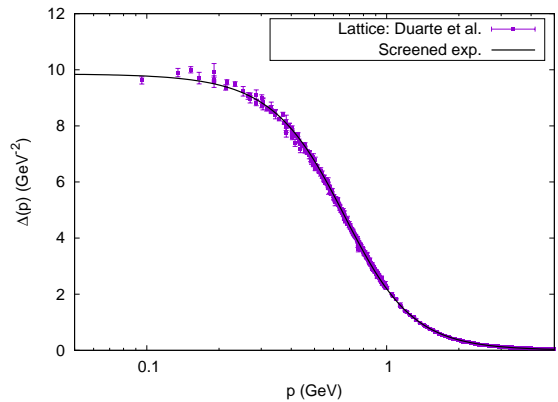}
\caption{Transverse dressed gluon propagator in the Landau gauge ($\xi=0$) and in Euclidean space, computed in the screened expansion of pure YM theory by optimizing the additive renormalization constant $F_{0}$ based on principles of gauge invariance. The lattice data are taken from Ref.~\cite{duarte}.}
\end{figure}

\section{The screened massive expansion of full QCD}

In this section we will extend the screened massive expansion to full QCD with one flavor of quarks. As we will see, our formalism is able to describe the non-perturbative generation of an infrared dynamical mass both for the chiral and the light quarks.

Our starting point is the formalism laid out in Sec.~IIA. After introducing the quarks in the Faddeev-Popov Lagrangian of pure Yang-Mills theory, we will perform a non-ordinary renormalization and split of the quark Lagrangian into a kinetic and an interaction term plus renormalization counterterms. The procedure parallels what we previously did for the gauge sector, but has a new feature, namely, the non-renormalization of the quark's bare mass. The motivation and consistency of such a choice will be discussed in Sec.~IIIA. In Sec.~IIIB we will define three resummation schemes for the dressed quark propagator, which differ by how the internal gluon line is treated in the quark self-energy.

\subsection{Setup and renormalization}

The Lagrangian of full QCD with one flavor of quarks is given by
\BE
\mc{L}_{\tx{QCD}}=\mc{L}+\mc{L}_{q,B},
\EE
where $\mc{L}$ is the Faddeev-Popov Lagrangian of pure Yang-Mills theory -- Eq.~\eqref{fplag} -- and $\mc{L}_{q,B}$ is the quark Lagrangian expressed in terms of the bare fields, mass and coupling,
\BE
\mc{L}_{\tx{q},B}=\psibar_{B}(i\slashed{D}_{B}-M_{B})\psi_{B}.
\EE
Here $M_{B}$ is the quark's bare mass, while $D_{B}$ is the bare covariant derivative acting on the bare quark field $\psi_{B}$,
\BE
D_{B}^{\mu}=\partial^{\mu}-ig_{B}A^{a\mu}_{B}T_{a}.
\EE

In order to renormalize the quark Lagrangian, we introduce a quark field renormalization constant $Z_{\psi}$ such that
\BE
\psi_{B}=Z_{\psi}^{1/2}\psi,
\EE
where $\psi$ is the renormalized quark field. Then $\mc{L}_{q,B}$ can be expressed as
\BE
\mc{L}_{\tx{q},B}=\psibar(i\slashed{D}-M_{B}Z_{\psi})\psi+\mc{L}_{q,\tx{c.t.}},
\EE
where $D$ is the renormalized covariant derivative acting on the renormalized quark field,
\BE
D_{\mu}=\partial_{\mu}-igA^{a}_{\mu}T_{a}
\EE
-- $g$ and $A_{\mu}^{a}$ being the renormalized coupling and gluon field defined as in Sec.~IIA --, while $\mc{L}_{q,\tx{c.t.}}$ contains the quark's field strength and quark-gluon vertex renormalization counterterms.

At this point, if the quark is not massless (i.e. $M_{B}\neq 0$), one usually introduces a renormalized quark mass through a kinetic term of the form $-M_{R}\,\psibar\psi$, and includes the corresponding mass renormalization counterterm $-\delta_{M}\,\psibar\psi$ into $\mc{L}_{q,\tx{c.t.}}$. In ordinary perturbation theory, $M_{R}$ and $M_{B}$ are understood to be proportional and related to each other by radiative corrections which can be computed perturbatively at any given loop order. Due to dynamical mass generation, however, in the IR the light quarks acquire a mass which is much larger than their renormalized mass $M_{R}$ and non-proportional to it; indeed, the former would be non-zero (and of the order of the QCD scale $\Lambda_{\tx{QCD}}$) also in the case of chiral quarks ($M_{B}=0$). Clearly, choosing $M_{R}$ as the mass of the zero-order propagator around which to expand the perturbative series, is not optimal for the purpose of exploring the low-energy dynamics of the quark sector.

On the other hand, the situation could improve if an effective mass scale, mimicking the dynamically generated IR quark mass, was used in place of the renormalized mass $M_{R}$. Our setup, therefore, employs the following scheme. As in ordinary perturbation theory, we add to the quark Lagrangian a mass term of the form $-M\,\psibar\psi$. However, we do not interpret $M$ as the renormalized counterpart of $M_{B}$. Instead, we regard the former as being a non-perturbative mass scale arising from the low-energy dynamics of the theory, and leave $M_{B}$ unrenormalized. Explicitly, we rewrite the quark Lagrangian as
\BE
\mc{L}_{\tx{q},B}=\psibar(i\slashed{D}-M)\psi+\psibar(M-M_{B}Z_{\psi})\psi+\mc{L}_{q,\tx{c.t.}}
\EE
and treat $M$ and $M_{B}$ as \textit{independent} mass parameters: the difference $M_{B}Z_{\psi}-M$, which in ordinary perturbation theory would correspond to the mass renormalization counterterm $\delta_{M}$, is not taken to be proportional to the coupling constant (i.e. small in the perturbative sense), nor is it regarded as fixed by the renormalization of the quark propagator. We anticipate that an appropriate choice of the diagrams to include in the one-loop quark propagator preserves the renormalizability of the theory also when using this non-standard scheme.

The quark Lagrangian is now split into a kinetic (zero-order) term $\mc{L}_{q,0}$,
\BE
\mc{L}_{q,0}=\psibar(i\slashed{\partial}-M)\psi,
\EE
in which $M$ appears as the mass in the zero-order quark propagator; an interaction term $\mc{L}_{q,\tx{int}}$,
\BE
\mc{L}_{q,\tx{int}}=\psibar(g\slashed{A}^{a}T_{a}+M-M_{B}Z_{\psi})\psi,
\EE
which contains the quark-gluon vertex and two new quadratic terms, proportional to $M$ and $M_{B}$; and a renormalization term $\mc{L}_{q,\tx{c.t.}}$,
\BE
\mc{L}_{q,\tx{c.t.}}=\psibar(i\delta_{\psi}\slashed{\partial}+g\,\delta_{g}\slashed{A}^{a}T_{a})\psi,
\EE
which contains the quark field strength renormalization counterterm $\delta_{\psi}=Z_{\psi}-1$ and a renormalization counterterm $\delta_{g}$ for the quark-gluon vertex.

The addition and subtraction of the mass term $-M\,\psibar\psi$ from the quark Lagrangian parallels what we did in the gluon sector of pure Yang-Mills theory. This is best seen in the chiral limit ($M_{B}\to 0$), where the addition of a mass term of the form $-M_{R}\,\psibar\psi$ would be meaningless, since $M_{R}\propto M_{B} = 0$. As a non-perturbative mass parameter not directly related to $M_{B}$, $M$ has the same status of the gluon mass parameter $m$ in the screened expansion of YM theory, and is allowed to remain finite also in the chiral limit. For this reason, we will refer to $M$ as the \textit{chiral mass} of the quark.

As in the screened expansion of YM theory, the shift of the quark Lagrangian changes the Feynman rules of the theory. First of all, the chiral mass $M$ now figures as the tree-level mass in the zero-order quark propagator $S_{M}(p)$,
\BE\label{propq1}
S_{M}(p)=\frac{i}{\slashed{p}-M}.
\EE
Second of all, two new two-point vertices $\delta\Gamma_{q,1}$ and $\delta\Gamma_{q,2}$ arise in the interaction:
\BE\label{qmcts}
\delta\Gamma_{q,1}(p)=iM,\quad\quad \delta\Gamma_{q,2}(p)=-iM_{B}Z_{\psi}.
\EE
We reiterate that in our framework these are treated as independent vertices. The quark-gluon interaction and renormalization vertices, on the other hand, are left unchanged, except for the quark mass renormalization counterterm, which must not be included in the calculation.

These Feynman rules must of course be supplied with those of the gluon sector, which were derived in Sec.~IIA in the context of pure YM theory. In particular, the transverse gluons propagate with a massive zero-order propagator -- Eq.~\eqref{glupropzero} --, and a third two-point vertex, the gluon mass counterterm of Eq.~\eqref{masscount}, is included in the interaction.

As a consequence of the new Feynman rules, the screened expansion of full QCD is non-perturbative in nature. Like in pure YM theory, this is due to the two-point vertices $\delta\Gamma_{g}$, $\delta\Gamma_{q,1}$ and $\delta\Gamma_{q,2}$, which are proportional to the gluon and the quark mass parameters $m^{2}$, $M$ and $M_{B}$, and are not taken to be proportional to the strong coupling constant.\\

Let us now turn our attention to how to compute the quark propagator in the new framework. The dressed quark propagator $S(p)$ can be expressed in terms of the 1PI quark self-energy $\Sigma(p)$ \footnote{Not to be confused with the ghost self-energy of Sec.~IIA.} as
\BE\label{Sp1}
S(p)=\frac{i}{\slashed{p}-M-\Sigma(p)}.
\EE
Due to the shift of the expansion point, $\Sigma(p)$ receives tree-level contributions not only from the quark field strength renormalization counterterm $\delta_{\psi}=Z_{\psi}-1$, but also from the new vertices $\delta\Gamma_{q,1}$ and $\delta\Gamma_{q,2}$ -- diagrams (1a) and (1b) in Fig.~4: we have
\BE\label{sigmaq1}
\Sigma(p)=-\delta_{\psi}\slashed{p}-M+M_{B}Z_{\psi}+\Sigma^{(\tx{loops})}(p),
\EE
where $\Sigma^{(\tx{loops})}(p)$ is the self-energy contribution coming from the loops. It follows that
\BE
[-iS(p)]^{-1}=Z_{\psi}\slashed{p}-M_{B}Z_{\psi}-\Sigma^{(\tx{loops})}(p).
\EE
As in pure YM theory, the mass $M$ introduced by the shift of the quark Lagrangian disappears from the propagator and the bare mass is restored at tree level, up to field-strength renormalization. In order to define the quark mass function $\mc{M}(p^{2})$ and $Z$-function $Z(p^{2})$, we first subdivide $\Sigma^{(\tx{loops})}(p)$ into a vector and a scalar term,
\BE\label{sigmaq2}
\Sigma^{(\tx{loops})}(p)=\slashed{p}\,\Sigma_{V}(p^{2})+\Sigma_{S}(p^{2}),
\EE
and then define two scalar functions $A(p^{2})$ and $B(p^{2})$,
\begin{align}\label{AandB}
\nn A(p^{2})&=Z_{\psi}-\Sigma_{V}(p^{2}),\\
B(p^{2})&=M_{B}Z_{\psi}+\Sigma_{S}(p^{2}).
\end{align}
In terms of $A(p^{2})$ and $B(p^{2})$, the functions $\mc{M}(p^{2})$ and $Z(p^{2})$ read
\begin{align}\label{ZandM}
Z(p^{2})=\frac{1}{A(p^{2})},\quad\quad \mc{M}(p^{2})=\frac{B(p^{2})}{A(p^{2})};
\end{align}
moreover, Eq.~\eqref{Sp1} can be rewritten as
\BE
S(p)=\frac{iZ(p^{2})}{\slashed{p}-\mc{M}(p^{2})}.
\EE

From Eqs.~\eqref{AandB} and \eqref{ZandM} we see that in the chiral limit ($M_{B}\to 0$), despite the absence of a tree-level mass for the quark propagator, the quark mass function $\mc{M}(p^{2})$ does not vanish: thanks to the finiteness of the non-perturbative scale $M$, one finds that $\Sigma_{S}(p^{2})\neq 0$, which makes $B(p^{2})\neq 0$ and thus $\mc{M}(p^{2})\neq 0$, also for vanishing $M_{B}$. Since $\Sigma_{S}(p^{2})$ comes from the loops, the mass of the quark is genuinely dynamical, a feature that was already highlighted in Sec.~II for the gluons in pure YM theory. For non-chiral quarks the situation is similar, the only difference being that $B(p^{2})$ also contains one additional tree-level term which is proportional to the bare mass $M_{B}$ of the quark. As we will see in a moment, the fact that this term is not renormalized poses no issue of consistency.

To one loop, an infinite number of diagrams contributes to the 1PI quark self-energy. These have the structure of the ordinary one-loop diagram of standard perturbation theory -- diagram (2a) in Fig.~4 --, with an arbitrarily large number of insertions of the gluon mass counterterm $\delta\Gamma_{g}$ and of the quark mass counterterms $\delta\Gamma_{q,1}$ and $\delta\Gamma_{q,2}$. In order to chose a truncation scheme for this infinite series, let us have a look at the first few such diagrams.

\begin{figure}[t]
\vskip 1cm
\centering
\includegraphics[width=0.45\textwidth]{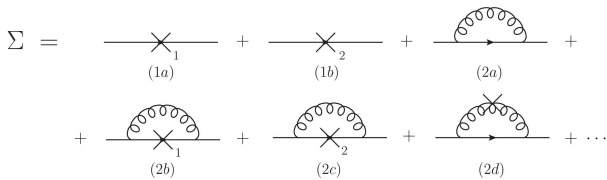}
\caption{1PI diagrams for the screened expansion one-loop quark self-energy. The crosses denote insertions of the mass counterterms. The subscripts 1 and 2 label the vertices $\delta\Gamma_{q,1}$ and $\delta\Gamma_{q,2}$ in Eq.~\eqref{qmcts}. The renormalization counterterms are not shown in the figure.}
\end{figure}

The simplest one-loop self-energy diagram is the ordinary uncrossed loop -- denoted by (2a) in Fig. 4. In a general covariant gauge, diagram (2a) has divergences in both its vector component and in its scalar component:
\BE\label{sig1div}
\Sigma^{(2a)}(p)=\left(c_{2aV}\,\slashed{p}+c_{2aS}\,M\right)\,\frac{2}{\epsilon}+\cdots,
\EE
where $c_{2aV}$ and $c_{2aS}$ are $O(g^{2})$ coefficients, $\epsilon=4-d$ is the regulator of dimensional regularization and the dots denote finite self-energy terms. While the vector divergence can be straightforwardly absorbed into the renormalization constant $Z_{\psi}$ -- see the first of Eq.~\eqref{AandB} --, in order to remove the mass divergence $c_{2aS}$ we would need to define a renormalized mass $M_{R}$ in terms of which
\BE
M_{B}=Z_{\psi}^{-1}\left(M_{R}-c_{2aS}\,M\,\frac{2}{\epsilon}+\tx{scheme-dep. consts.}\right)
\EE
-- see the second of Eq.~\eqref{AandB}. A relation like this mixes infrared entities (namely, the chiral mass $M$) to UV features (the divergence and the renormalization of the bare mass) with no apparent logic, aside from the mathematical convenience of it. Moreover, this type of renormalization cannot be employed in the chiral limit $M_{B}\to 0$, when there is no bare mass in which to absorb the divergence. For these reasons, it must be rejected.

We note that, having been introduced through a term which is added and subtracted in the Lagrangian, the mass parameter $M$ cancels in the total action; as a consequence, any divergence proportional to $M$ must disappear when diagrams with a different number of mass counterterms are resummed at the same loop order.

In fact, diagram (2b) in Fig.~4 is easily shown to contain the same mass divergence of diagram (2a) with an opposite sign: since the crossed quark line in the diagram can be expressed as a derivative with respect to the quark's chiral mass,
\BE
\frac{i}{\slashed{p}-M}(iM)\frac{i}{\slashed{p}-M}=-M\frac{\partial}{\partial M}\,\frac{i}{\slashed{p}-M},
\EE
the self-energy contribution from diagram (2b), $\Sigma^{(2b)}(p)$, can be obtained as a derivative of $\Sigma^{(2a)}(p)$:
\BE\label{der1}
\Sigma^{(2b)}(p)=-M\frac{\partial}{\partial M}\,\Sigma^{(2a)}(p).
\EE
It follows that
\BE
\Sigma^{(2b)}(p)=-c_{2aS}\,M\,\frac{2}{\epsilon}+\cdots,
\EE
that is, $\Sigma^{(2a)}(p)$ and $\Sigma^{(2b)}(p)$ have opposite mass divergences. As a consequence, the sum of diagrams (2a) and (2b) only contains a divergence in the vector component, coming entirely from $\Sigma^{(2a)}(p)$. This divergence can be shown to be the same as the one found in ordinary perturbation theory, and is to be absorbed into the definition of $Z_{\psi}$, as we saw earlier.

Now, in the Landau gauge ($\xi=0$), the divergence contained in $\Sigma^{(2a)}(p)$ is known from ordinary perturbation theory to vanish. Therefore, not only does the sum $\Sigma^{(2a)}(p)+\Sigma^{(2b)}(p)$ not contain mass divergences, but in the Landau gauge it is also fully finite. In particular, if we truncate the perturbative series to diagrams (2a) and (2b) in Fig.~4 and limit ourselves to the Landau gauge, then the term $M_{B}Z_{\psi}$ that appears in the $B(p^{2})$ function -- see Eq.~\eqref{AandB} -- can be taken to be a finite constant. In other words, no renormalization of divergent constants or masses is required in the screened expansion of the Landau gauge quark propagator, provided that the latter is truncated to diagrams (2a) and (2b).

On the other hand, if $\xi\neq 0$, the vector divergence in $\Sigma^{(2a)}(p)+\Sigma^{(2b)}(p)$ still needs to be absorbed into $Z_{\psi}$. For non-chiral quarks ($M_{B}\neq 0$), if $M_{B}$ were taken to be finite, this would leave us with a divergent $M_{B}Z_{\psi}$ term inside $B(p^{2})$. Therefore, for $\xi\neq 0$ and $M_{B}\neq 0$, a renormalized mass $M_{R}$ must still be introduced, even when truncating the quark self-energy to diagrams (2a) and (2b).

It is easy to see that a renormalized mass of the form $M_{R}=M_{B}Z_{\psi}$ would not have the ordinary behavior of a running mass under the renormalization group (RG). Indeed, if the RG equations were employed in the scheme, $M_{R}$ would run exclusively with the anomalous dimension of the quark field, rather than with the full anomalous dimension of the quark mass. This happens because we have left out one further divergent diagram from the calculation, namely, diagram (2c) in Fig.~4. The latter can be obtained from diagram~(2a) by using the equality
\BE
\frac{i}{\slashed{p}-M}(-iM_{B}Z_{\psi})\frac{i}{\slashed{p}-M}=M_{B}Z_{\psi}\frac{\partial}{\partial M}\,\frac{i}{\slashed{p}-M},
\EE
which can be exploited to write
\BE\label{der2}
\Sigma^{(2c)}(p)=M_{B}Z_{\psi}\frac{\partial}{\partial M}\,\Sigma^{(2a)}(p).
\EE
In particular,
\BE
\Sigma^{(2c)}(p)=c_{2aS}\,M_{B}Z_{\psi}\,\frac{2}{\epsilon}+\cdots.
\EE
As we can see, diagram (2c) has a scalar divergence proportional to $M_{B}Z_{\psi}$. When the latter is summed to the tree-level term in $B(p^{2})$, one finds
\BE
B(p^{2})=M_{B}Z_{\psi}\left(1+c_{2aS}\,\frac{2}{\epsilon}\right)+\cdots.
\EE
By simple dimensional arguments, it is easy to show that the remaining one-loop diagrams in the quark self-energy are finite. Therefore, the above expression spells out the complete divergent term of the scalar component of the one-loop self-energy, obtained by summing the divergences of diagrams (2a) to (2c) in Fig.~4. Such a term can indeed be equated, modulo finite constants, to a renormalized mass $M_{R}$ which would run like an ordinary quark mass if the RG equations were to be used, leaving us with
\BE
B(p^{2})=M_{R}+\tx{finite terms}.
\EE
Beyond the Landau gauge, then, consistency with the renormalization group requires us to include diagram (2c) in the calculation. In the Landau gauge, on the other hand, diagram (2c) is not needed in principle, since to one loop the sum of diagrams (2a) and (2b) already results in a finite quark 1PI self-energy.

Despite being necessary for theoretical consistency, if the renormalized quark mass $M_{R}$ is much smaller than the chiral mass $M$, the inclusion of diagram (2c) in the quark self-energy turns out not to be essential from a quantitative point of view. This is easily seen as follows. Let $\Sigma^{(2a,2b,2c)}_{f}(p)$ be the finite parts of the self-energy diagrams (2a), (2b) and (2c). Using Eqs.~\eqref{der1} and \eqref{der2},
\BE
\Sigma^{(2b)}_{f}(p)+\Sigma^{(2c)}_{f}(p)=-(M-M_{B}Z_{\psi})\,\frac{\partial}{\partial M}\,\Sigma^{(2a)}_{f}(p).
\EE
Modulo higher-order corrections, we can set $M_{B}= M_{R}$, $Z_{\psi}= 1$ in the above equation, so that
\BE
\Sigma^{(2b)}_{f}(p)+\Sigma^{(2c)}_{f}(p)=-(M-M_{R})\,\frac{\partial}{\partial M}\,\Sigma^{(2a)}_{f}(p).
\EE
It is then clear that, as long as $M_{R}\ll M$, the contribution of diagram (2c) is completely negligible with respect to that of diagram (2b). In other words, for the light quarks, diagram (2c) can be taken to contribute only to the divergent part of the self-energy, i.e. to the renormalization of the bare mass\footnote{For the sake of completeness, we note that there is one catch in this argument: at high energies, the scalar part of the sum of diagrams (2a) and (2b) can be shown to vanish -- see e.g. Sec.~IVA~--, so that, instead of being negligible, diagram (2c) actually makes up for the whole scalar self-energy. As long as we limit ourselves to low and moderate energies, this issue does not arise. At large energies, however, diagram (2c) and appropriate RG techniques are needed to account for the correct asymptotic behavior of the quark mass function.}.

To summarize, in every linear covariant gauge, diagram (2b) in Fig.~4 is needed in order to remove the mass divergence in diagram (2a). This mass divergence has no counterpart in ordinary perturbation theory, since it is proportional to the quark chiral mass $M$. Diagram (2c) is essential to renormalize the bare mass $M_{B}$ in compliance with the standard RG equations. Nonetheless, its finite part is completely negligible in the case of light quarks. Finally, in the Landau gauge the sum of diagrams (2a) and (2b) results in a finite self-energy. Since for the light quarks diagram (2c) is quantitatively negligible, in the Landau gauge one can simply exclude it from the self-energy and interpret the free parameters $M_{B}$ and $Z_{\psi}$ as bare but finite quantities.\\

In the next section we will carry on with our analysis of the resummation of the one-loop quark propagator. Our main focus will be on exploring different ways to treat the finite diagrams in Fig.~4.

\subsection{Resummation schemes for the quark propagator}

Up to this point we have discussed the self-energy diagrams which contribute to the divergent part of the one-loop quark propagator, namely, diagrams (2a) to (2c) in Fig.~4. Using simple dimensional arguments, it is easy to show that, to one loop, other insertions of the gluon and quark two-point mass counterterms indeed yield convergent diagrams. As an example, consider diagram (2d) in Fig.~4. This diagram has a superficial degree of divergence $D$
\BE
D=d-1-2-2\to -1<0
\EE
-- where the $-1$ and the $-2$'s come from the internal quark and gluon lines, respectively --, making diagram (2d) UV-finite in the limit $d\to 4$. Equivalently, observe that diagram (2d) can be expressed as a derivative of diagram (2a) with respect to the gluon mass parameter $m^{2}$: using Eq.~\eqref{gluonders} with $n=1$ we find that
\BE
\Sigma^{(2d)}(p)=-m^{2}\frac{\partial}{\partial m^{2}}\,\Sigma^{(2a)}(p).
\EE
Since the divergent part of $\Sigma^{(2a)}(p)$ does not depend on $m^{2}$, $\Sigma^{(2d)}(p)$ is again shown to be finite.

While divergent diagrams are included in the one-loop calculation based on principles of renormalizability, assessing which finite diagrams should be included as well is far more tricky. One option could be to adopt a minimalistic point of view and limit oneself to the one-loop diagrams needed for consistency, i.e. diagrams (2a) to (2b) or (2c) in Fig.~4. Yet another option could be to retain all the one-loop diagrams with a maximum of three vertices, as we did for the gluon propagator in Sec.~II; in practice, this amounts to also including diagram (2d) in the self-energy. These two resummation schemes differ by how the internal gluon line is treated -- explicitly, by whether the internal zero-order gluon propagator is corrected with its own mass counterterm or not. We will refer to them as the minimalistic and the vertex-wise schemes, respectively. Schemes with a larger number of crossed diagrams (not shown in Fig.~4) are not considered in this paper.

\begin{figure}[b]
\vskip 1cm
\centering
\includegraphics[width=0.25\textwidth]{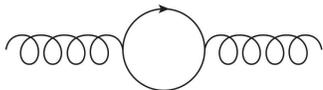}
\caption{Quark loop in the unquenched gluon polarization. To one loop, its inclusion affects the value of the gluon mass parameter $m^{2}$ and the position and residue of the poles of the gluon propagator.}
\end{figure}

In the next section we will fit and compare the results obtained in the minimalistic and vertex-wise schemes with the quenched lattice data of Ref.~\cite{kamleh}. The reason for using quenched rather than unquenched lattice data is to exploit our previous results for pure YM theory and fix ab initio the value of the gluon mass parameter $m^{2}$ that appears in the quark propagator -- thus reducing the number of free parameters to be fitted. Indeed, observe that, to one loop, the quark self-energy diagrams for the quenched and unquenched theories coincide. Hence, in principle, our results could be used for comparisons with both quenched and unquenched data. However, in the framework of the screened massive expansion, the value of the gluon mass parameter $m^{2}$ running in diagrams (2a)-(2d) (Fig.~4) can receive corrections from the quark loop in the gluon polarization (Fig.~5), which is only present in the unquenched theory. Thus we expect the value of $m^{2}$ to be different depending on which theory (quenched or unquenched) we are trying to fit. In order to reduce the freedom in the choice of free parameters, we decide not to make a new determination of the gluon mass parameter, but rather to use the quenched lattice data for our fits. The value $m=0.6557$~GeV was obtained in~\cite{xigauge} by a fit of the lattice data of Ref.~\cite{duarte} for pure YM theory. With $m$ fixed, the remaining free parameters of the quark propagator are the chiral mass $M$, the quark bare mass $M_{B}$ or renormalized mass $M_{R}$, and the renormalization constants.

As we will see, the minimalistic and vertex-wise schemes are practically equivalent from the point of view of the fit, the only difference being in the values of the parameters needed to achieve the match with the lattice data. Both of them succeed in quantitatively reproducing the lattice mass function $\mc{M}(p^{2})$ with very good precision. On the other hand, in none of the two the $Z$-function has the behavior displayed by the lattice data: $Z(p^{2})$ is found to be a decreasing function of momentum, at variance with the lattice. To one loop, such a mismatch is not unseen, having been reported for another massive model, namely the Curci-Ferrari model of Ref.~\cite{tissier14}.\\

One interesting question to ask is whether higher-order or non-perturbative corrections to the internal gluon line in the quark self-energy can sensibly change the behavior of the $Z$-function. Indeed, as we noted in the Introduction, in the Landau gauge, to one loop and at sufficiently high energies, $Z(p^{2})\approx1$, making the $Z$-function sensitive to all kinds of contributions beyond the leading perturbative order. The near vanishing of the perturbative contribution makes the $Z$-function a valid benchmark for investigating the role of condensates by the OPE. Indeed, the slightly increasing behavior which is observed on the lattice has been modeled by OPE \cite{arriola,blossier,wang} and shown to be consistent with the existence of a dimension-$2$ gluon condensate of the form $\avg{A^{2}}$. In order to explore these issues, we introduce a third resummation scheme, which we term the complex-conjugate (CC) scheme for reasons that will become apparent in a moment.

In the CC scheme, instead of only summing the zero-order gluon propagator (minimalistic scheme) or its counterterm-corrected counterpart (vertex-wise scheme), we use the fully dressed gluon propagator as the internal gluon line of the one-loop quark self-energy (see Fig.~6). Switching to the dressed gluon propagator allows us to account for the full non-perturbative dynamics of the gluon, when computing the quark propagator.

While in principle using the dressed propagator would require us to resum and integrate an infinite number of higher-order diagrams, in practice we know that -- in pure Yang-Mills theory -- the principal part of the screened expansion's one-loop gluon propagator provides a very good approximation to the dressed propagator, modulo a multiplicative factor (see Sec.~IIB, in particular Figs.~2 and 3). Therefore, in the CC scheme, we use a zero-order gluon propagator which -- in Euclidean space and in the Landau gauge -- reads
\BE
\Delta^{(\tx{c.c.})}_{\mu\nu}(p)=\left\{\frac{R}{p^{2}+p_{0}^{2}}+\frac{\overline{R}}{p^{2}+\overline{p_{0}^{2}}}\right\}\ t_{\mu\nu}(p).
\EE
Here $p_{0}^{2}$ and $\overline{p_{0}^{2}}$ are the complex-conjugate poles of the dressed gluon propagator (hence the name CC scheme) in the complexified Minkowski space, and $R$ and $\overline{R}$ are their normalized residues. The value of the modulus $|R|$ -- which depends both on the renormalization conventions for the dressed gluon propagator and on a multiplicative factor that converts between the full propagator and its principal part -- does not actually affect the results for the quark propagator, provided that the free parameters are suitably redefined. Indeed, to one loop, the internal gluon line in the quark self-energy is multiplied by a factor of the strong coupling constant $\alpha_{s}$, so that $|R|$ can be absorbed into the definition of the latter.  Our convention for the definition of $|R|$ (and thus also $\alpha_{s}$ in the CC scheme) will be discussed in Sec.~IVC. As for $p_{0}^{2}$ and the phase of $R$, we use the values reported in Tab.~I (Sec.~IIB). These were obtained in pure Yang-Mills theory and are thus suitable for calculations in the quenched theory, in line with our discussion on the gluon mass parameter $m^{2}$ in the minimalistic and vertex-wise schemes.

\begin{figure}[t]
\vskip 1cm
\centering
\includegraphics[width=0.45\textwidth]{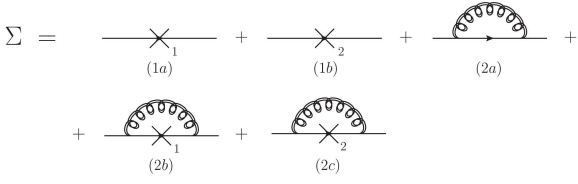}
\caption{1PI diagrams for the quark self-energy in the complex-conjugate (CC) scheme. The double lines represent the fully dressed gluon propagator, which in the CC scheme is approximated by the principal part of the one-loop gluon propagator (Sec.~II).}
\end{figure}

As we show in Appendix~B, despite the poles $p_{0}^{2}$ and $\overline{p_{0}^{2}}$ being complex, as long as the external momentum $p^{2}\in \Bbb{R}$, the loop integrals in the CC scheme can be computed by employing the usual machinery of Feynman parameter integrals and Gamma functions. In particular, if we denote with $\Sigma^{(\tx{loops})}_{\tx{m.}}(p)$ the loop contribution to quark self-energy computed in the minimalistic scheme -- diagrams (2a) to (2c) in Fig.~4 --, then we can express the corresponding self-energy term  $\Sigma^{(\tx{loops})}_{\tx{c.c.}}(p)$ in the CC scheme as
\begin{align}\label{ccms}
\nn&\Sigma^{(\tx{loops})}_{\tx{c.c.}}(p)=\\
&=R\,\Sigma^{(\tx{loops})}_{\tx{m.}}(p)\Big|_{m^{2}=p_{0}^{2}}+\overline{R}\,\Sigma^{(\tx{loops})}_{\tx{m.}}(p)\Big|_{m^{2}=\overline{p_{0}^{2}}}
\end{align}
or equivalently
\BE
\Sigma^{(\tx{loops})}_{\tx{c.c.}}(p)=2\tx{Re}\left\{R\,\Sigma^{(\tx{loops})}_{\tx{m.}}(p)\Big|_{m^{2}=p_{0}^{2}}\right\}.
\EE

As we will see, the $Z$-function computed in the CC scheme indeed turns out to have a qualitatively different behavior than those computed in the minimalistic or vertex-wise scheme, closer to the one displayed by the quenched lattice data at moderately large momenta.

\section{The quark propagator in the Landau gauge}

In this section we report our results for the quark propagator in the Landau gauge using the screened massive expansion of full QCD in the minimalistic, vertex-wise and complex-conjugate resummation schemes introduced in Sec.~IIIB. As previously discussed, we will use the lattice data of Ref.~\cite{kamleh} for quenched QCD in order to test the validity of the expansion and fit the free parameters that appear in the propagator. These parameters are defined in what follows.\\

In general -- see Eqs.~\eqref{AandB} and \eqref{ZandM} --, the quark mass and $Z$-function can be expressed as
\begin{align}\label{MandZ2}
\nn\mc{M}(p^{2})&=\frac{M_{B}Z_{\psi}+\Sigma_{S}(p^{2})}{Z_{\psi}-\Sigma_{V}(p^{2})},\\
Z(p^{2})&=[Z_{\psi}-\Sigma_{V}(p^{2})]^{-1}.
\end{align}
Here $\Sigma_{V}(p^{2})$ and $\Sigma_{S}(p^{2})$ are the vector and scalar components of the loop contribution to the quark self-energy, $M_{B}$ is the quark bare mass and $Z_{\psi}$ is the quark field renormalization constant. In the Landau gauge and to one loop, as we saw in Sec.~III, $\Sigma_{V}(p^{2})$ is UV-convergent. As a consequence, we can write
\BE\label{SVf}
\Sigma_{V}(p^{2})=\frac{\alpha_{s}}{3\pi}\ \sigma_{V}(p^{2}),
\EE
where $\sigma_{V}(p^{2})$ is a finite function. Nonetheless, the value of $Z_{\psi}$ still needs to be fixed. We decide to do so by renormalizing the $Z$-function in the momentum-subtraction (MOM) scheme at a specified renormalization scale $\mu^{2}$. Namely, we set
\BE\label{Zpsiren0}
Z(\mu^{2})=1\qquad\Longleftrightarrow\qquad Z_{\psi}-\Sigma_{V}(\mu^{2})=1,
\EE
or, equivalently,
\BE\label{Zpsiren}
Z_{\psi}=1+\frac{\alpha_{s}}{3\pi}\,\sigma_{V}(\mu^{2}),
\EE
where we take $\mu$ to be equal to 4~GeV. As we will see in a moment, as far as the fits are concerned, this choice is inessential to our results.

At variance with $\Sigma_{V}(p^{2})$, the scalar component $\Sigma_{S}(p^{2})$ can be either UV-divergent or UV-convergent depending on whether diagram (2c) in Figs.~4 and 6 is included or not in the self-energy, respectively. In the absence of diagram (2c), $\Sigma_{S}(p^{2})$ can be expressed as
\BE\label{SSf}
\Sigma_{S}(p^{2})=\frac{\alpha_{s}}{\pi}\ \sigma_{S}(p^{2}),
\EE
where $\sigma_{S}(p^{2})$ is a finite function. In particular, it follows from the first of Eq.~\eqref{MandZ2} that $M_{B}$ must be taken to be finite. If we now define two finite constants $h_{0}$ and $k_{0}$,
\begin{align}\label{h0k0def}
\nn h_{0}&=\frac{3\pi}{\alpha_{s}}Z_{\psi},\\
k_{0}&=\frac{\pi}{\alpha_{s}}\, M_{B}Z_{\psi},
\end{align}
then the mass function $\mc{M}(p^{2})$ reads
\begin{align}\label{massk0h0}
\mc{M}(p^{2})=\frac{3[k_{0}+\sigma_{S}(p^{2})]}{h_{0}-\sigma_{V}(p^{2})};
\end{align}
here $\alpha_{s}$ and $M_{B}$ have been absorbed into the definition of $h_{0}$ and $k_{0}$.

While the exact propagator should not depend on the scale $\mu$, apart from a renormalization factor, the approximate one-loop function $\mc{M}(p^{2})$ still has an implicit spurious dependence on $\mu$ through the parameters $h_0$, $k_0$, according to Eqs.~\eqref{h0k0def} and \eqref{Zpsiren}. Thus, the one-loop result can be optimized by a wise choice of the parameters: fixing $h_{0}$ and $k_{0}$ amounts to choosing an optimal renormalization -- together with the corresponding coupling and bare mass -- for the quark mass function.

As discussed in Sec.~IIB, for the gluon propagator such an optimization can be achieved from first principles in pure YM theory. Here, we just assume the existence of an optimal value of the parameters and determine them by a comparison with the lattice data. Thus, $h_0$ and $k_0$ are regarded as free parameters which depend on the scale ambiguity of the loop expansion.\\

For our fits, we will use $h_{0}$, $k_{0}$ and the chiral mass $M$ as the primary free parameters. It follows that our choice of the MOM scheme with $\mu=4$~GeV as the renormalization scale has no impact on the results of the fit. What the renormalization scheme actually determines is the value of $\alpha_{s}$, which can be computed at fixed $h_{0}$ and $M$ by using Eq.~\eqref{Zpsiren} and the first of Eq.~\eqref{h0k0def}:
\BE
\alpha_{s}=3\pi\left[h_{0}-\sigma_{V}(\mu^{2})\right]^{-1}.
\EE
From the above equation, $\alpha_{s}$ could be interpreted as the strong coupling constant defined at the renormalization scale $\mu=4$~GeV. However, it must be kept in mind that the renormalization prescription we chose is fully arbitrary. Actually, if the $Z$-function computed in the screened expansion is not well-behaved -- which is the case here, as we have anticipated --, then taking $Z(\mu^{2})=1$ as the starting point for measuring $\alpha_{s}$ could lead to meaningless values for the coupling constant. For the same reason, while in principle the lattice data for the $Z$-function could be used to fit at least some of the parameters of the expansion, we will instead fully rely on the lattice data for the quark mass function to perform the fit.

For completeness, we will also report our results in terms of the renormalized mass $M_{R}$. As we saw in Sec.~IIIA, the latter must be introduced as soon as diagram (2c) is included in the quark self-energy. This is due to the fact that, in the presence of said diagram, $\Sigma_{S}(p^{2})$ contains a divergence proportional to $M_{B}Z_{\psi}$. Namely, for $N=3$, in the minimalistic and vertex-wise schemes\footnote{For the complex-conjugate scheme see ahead, Sec.~IVC.},
\BE\label{SS2c}
\Sigma_{S}(p^{2})=\frac{\alpha_{s}}{\pi}\ \left[\sigma_{S}(p^{2})+M_{B}Z_{\psi}\,\frac{2}{\epsilon}\right].
\EE
Since, when $M_{R}\ll M$, the finite part of diagram (2c) is negligible -- see the discussion in Sec.~III --, the function $\sigma_{S}(p^{2})$ in Eq.~\eqref{SS2c} can be taken to be very same as the one in Eq.~\eqref{SSf}\footnote{The same goes for Eq.~\eqref{SVf}: $\Sigma_{V}(p^{2})$ is the same function both in the presence and in the absence of diagram (2c), with $\sigma_{V}(p^{2})$ unchanged.}. A renormalized mass $M_{R}$ can then be defined by absorbing the mass divergence of diagram (2c) into $M_{B}$,
\BE\label{MBren}
M_{R}=M_{B}Z_{\psi}\left(1+\frac{\alpha_{s}}{\pi}\,\frac{2}{\epsilon}\right).
\EE
With $M_{R}$ as above, Eq.~\eqref{massk0h0} still holds in the presence of diagram (2c), with the constant $k_{0}$ defined as
\BE\label{k02}
k_{0}=\frac{\pi}{\alpha_{s}}\,M_{R}
\EE
and $h_{0}$ defined in the first of Eq.~\eqref{h0k0def}. Of course, whether we express our results in terms of $M_{B}$ or of $M_{R}$ has no quantitative impact on our fits, since these are performed using $h_{0}$ and $k_{0}$, which as free parameters are more general than the masses and coupling themselves.\\

In the next sections, our focus will be on quarks whose lattice masses $M_{\tx{lat}} = 18,36,54,72,90$~MeV are small with respect to the QCD scale. Nonetheless, we will also present some results for heavier quarks.
 
\subsection{Minimalistic scheme}

\begin{table}[t]
\setlength{\tabcolsep}{10pt}
\def\arraystretch{1.5}
\begin{tabular}{c||c|c|c}
$M_{\tx{lat}}$&$M$&$h_{0}$&$k_{0}$\\
\hline
\hline
$18$&$368.6$&$2.132$&$-10.3$\\
$18^{\star}$&$318.1$&$1.791$&$6.0$\\
$36$&$330.8$&$1.967$&$14.1$\\
$54$&$320.0$&$2.073$&$38.1$\\
$72$&$330.7$&$2.341$&$62.4$\\
$90$&$336.9$&$2.504$&$88.6$
\end{tabular}
\caption{Fit parameters for the quark mass function $\mc{M}(p^{2})$ in the minimalistic scheme. $M_{\tx{lat}}$, $M$ and $k_{0}$ are expressed in MeV. The lattice data are taken from Ref.~\cite{kamleh}. The asterisked row was obtained at fixed $M_{B}$, see Tab.~III.}
\end{table}
\begin{table}[t]
\setlength{\tabcolsep}{10pt}
\def\arraystretch{1.5}
\begin{tabular}{c||c|c|c|c}
$M_{\tx{lat}}$&$M$&$\alpha_{s}$&$M_{B}$&$M_{R}$\\
\hline
\hline
$18$&$368.6$&$3.139$&$-14.4$&$-10.2$\\
$18^{\star}$&$318.1$&$3.542$&$10$&$6.7$\\
$36$&$330.8$&$3.322$&$21.5$&$14.9$\\
$54$&$320.0$&$3.202$&$55.2$&$38.9$\\
$72$&$330.7$&$2.935$&$79.9$&$58.3$\\
$90$&$336.9$&$2.793$&$106.1$&$78.8$
\end{tabular}
\caption{Fit parameters for the quark mass function $\mc{M}(p^{2})$ in the minimalistic scheme, in terms of $\alpha_{s}$ and $M_{B}$ or $M_{R}$ (renormalization scale: $\mu=4$~GeV). $M_{\tx{lat}}$, $M$, $M_{B}$ and $M_{R}$ are expressed in MeV. The lattice data are taken from Ref.~\cite{kamleh}. The asterisked row was obtained at fixed $M_{B}$.}
\end{table}

In the minimalistic resummation scheme, the loop diagrams included in the quark self-energy are those denoted by (2a), (2b) and, for the purpose of defining a renormalized mass $M_{R}$, (2c) in Fig.~4. The quark mass function $\mc{M}(p^{2})$ can be expressed as
\begin{align}\label{massk0h0ms}
\mc{M}(p^{2})=\frac{3[k_{0}+\sigma_{S}^{(\tx{m.})}(p^{2})]}{h_{0}-\sigma_{V}^{(\tx{m.})}(p^{2})},
\end{align}
where the analytic expressions for the scalar functions $\sigma_{S}^{(\tx{m.})}(p^{2})$ and $\sigma_{V}^{(\tx{m.})}(p^{2})$ are reported in Appendix~A. By fixing $m=655.7$~MeV as discussed in Sec.~IIIB and fitting the quenched lattice mass functions of Ref.~\cite{kamleh} for the lattice masses $M_{\tx{lat}} = 18,36,54,72,90$~MeV, we obtained the values of $h_{0}$ and $k_{0}$ reported in Tab.~II. In Tab.~III we list the corresponding values of $\alpha_{s}$, $M_{B}$ and $M_{R}$, computed by employing the definitions in Eqs.~\eqref{Zpsiren}, \eqref{h0k0def} and \eqref{k02}.

As we can see from Fig.~7, the mass functions computed in the minimalistic scheme show a very good agreement with the lattice data. For all but one of the considered lattice masses -- namely, $M_{\tx{lat}} = 18$~MeV, which we will discuss separately in a moment --, the fitted values of the chiral mass $M$ are found to be in the range $320-337$~MeV, while the bare masses $M_{B}$ are found to increase with $M_{\tx{lat}}$, always keeping close to the latter.

\begin{figure}[t]
\centering
\includegraphics[width=0.30\textwidth,angle=-90]{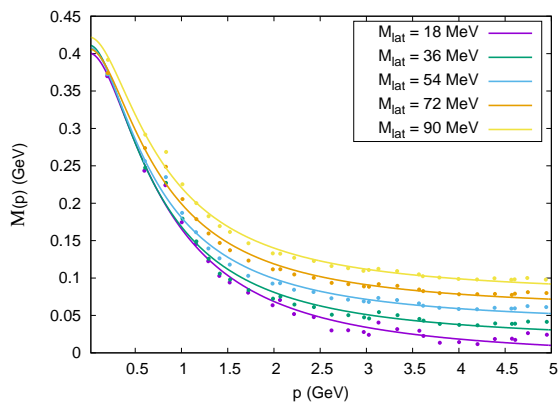}
\caption{Quark mass function $\mc{M}(p^{2})$ in the Euclidean space and in the Landau gauge for different values of the lattice mass $M_{\tx{lat}}$. Points: quenched lattice data from Ref.~\cite{kamleh}. Curves: one-loop mass functions computed in the minimalistic resummation scheme using the parameters in Tab.~II (equivalently, Tab.~III).}
\end{figure}

The fact that $M_{B}\approx M_{\tx{lat}}$ can be easily explained by looking at the high-momentum limit of the functions $\sigma_{V}^{\tx{(m.)}}(p^{2})$ and $\sigma_{S}^{\tx{(m.)}}(p^{2})$. For $p^{2}\gg m^{2},M^{2}$ we have
\begin{align}\label{UVA}
\nn\sigma_{V}^{\tx{(m.)}}(p^{2})&\to-1-\frac{3m^{2}}{4p^{2}}+\frac{3m^{2}}{2p^{2}}\ln\frac{p^{2}}{m^{2}}\to-1,\\
\sigma_{S}^{\tx{(m.)}}(p^{2})&\to\frac{2M^{2}}{p^{2}}\ln\frac{p^{2}}{M^{2}}\to0.
\end{align}
Therefore, in terms of $M_{B}$ and $\alpha_{s}$,
\BE\label{UVB}
\mc{M}(p^{2})\to\frac{M_{B}Z_{\psi}}{Z_{\psi}+\frac{\alpha_{s}}{3\pi}}\approx M_{B}\qquad (p^{2}\gg m^{2},M^{2}),
\EE
where the approximation holds provided that the coupling is sufficiently small. The above equation shows that the scale of the high-momentum limit of the mass function is set by the bare mass $M_{B}$; since on the lattice the same role is played by the lattice mass $M_{\tx{lat}}$, we expect $M_{B}\approx M_{\tx{lat}}$ as long as our function fits well the lattice data.

In the limit of vanishing momenta, regardless of the lattice mass, the data saturate to a finite value of about $350-450$~MeV\footnote{Note that this value is larger for the heavy quarks, as we will show later on in Fig.~10.}. The approximate independence of the saturation value from $M_{\tx{lat}}$ is expected on the basis that, in the infrared, the light quarks acquire most of their mass through the strong interactions, whose scale is much larger than the quark mass contained in the Lagrangian, and thus dominates over the latter. The mass function computed in the minimalistic scheme does reproduce this feature, provided that the chiral mass $M$ is comparable in value for the lattice masses under consideration (as is the case in our fits).
\begin{table}[t]
\setlength{\tabcolsep}{10pt}
\def\arraystretch{1.5}
\begin{tabular}{c|c||c|c|c}
$M_{\tx{lat}}$&$M_{B}$&$M$&$\alpha_{s}$&$M_{R}$\\
\hline
\hline
$18$&$0$&$338.1$&$3.373$&$0.0$\\
$18$&$10$&$318.1$&$3.542$&$6.7$\\
$18$&$18$&$302.7$&$3.679$&$11.9$
\end{tabular}
\caption{Fit parameters for the quark mass function $\mc{M}(p^{2})$ in the minimalistic scheme, in terms of $\alpha_{s}$ and $M_{B}$ or $M_{R}$ (renormalization scale: $\mu=4$~GeV), given $M_{\tx{lat}} = 18$~MeV and $M_{B}$ fixed to three different values. $M_{\tx{lat}}$, $M_{B}$, $M$ and $M_{R}$ are expressed in MeV. The lattice data are taken from Ref.~\cite{kamleh}.}
\end{table}
\begin{figure}[t]
\centering
\includegraphics[width=0.30\textwidth,angle=-90]{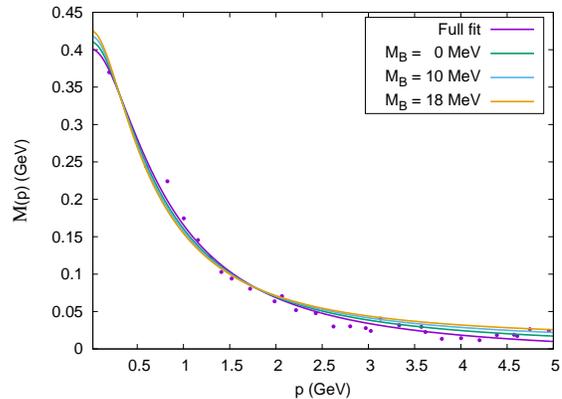}
\caption{$M_{\tx{lat}}=18$~MeV quark mass function in the Euclidean space and in the Landau gauge. Points: quenched lattice data from Ref.~\cite{kamleh}. Curves: one-loop mass functions computed in the minimalistic resummation scheme. The parameters for the curves with $M_{B}=0,10,18$~MeV are reported in Tab.~IV; those for the curve labeled as ``full fit'' are reported in Tab.~III.}
\end{figure}

In Tab.~III the value of the bare mass $M_{B}$ fitted for $M_{\tx{lat}}=18$~MeV stands out for being negative (this is a direct consequence of $k_{0}<0$ in Tab.~II). Presumably, this physically meaningless result is an artifact of the fit caused by the highly oscillatory tail of the $M_{B}=18$~MeV lattice mass function; the oscillations themselves are most likely due to discretization errors, as suggested by the large error bars in the original data (see Ref.~\cite{kamleh}). A constrained fit forcing $M_{B}\geq 0$ is not able to fix this issue, since, in the presence of the constraint, the fitting routine still tries to push $M_{B}$ to negative values, which implies that the lower boundary of the fitting interval -- namely $M_{B}=0$ -- is inevitably hit. Thus no meaningful result for $M_{B}$ is obtained by constraining the latter to be non-negative. Cutting the data at large momenta in order to avoid the oscillations (which begin at approximately $2.5-3$~GeV), as well, would not improve the situation: since at low momenta the quark mass function is not very sensitive to the value of $M_{B}$ (provided, of course, that we assume $M_{B}\ll M$), employing a cut dataset would make it impossible to meaningfully establish the value of the bare mass by a fit. As an alternative, to test our results, we checked that fixing the value of $M_{B}$ by hand, instead of fitting it from the lattice data, still yields a mass function which -- modulo oscillations -- is in good agreement with the lattice. Some examples are shown in Fig.~8, where we plot the data for $M_{\tx{lat}}=18$~MeV together with our minimalistic scheme mass function. Here $M_{B}$ is set to $0,10,18$~MeV, while the rest of the free parameters (reported in Tab.~IV) are still obtained by fitting the data. Remarkably, as soon as the bare mass is fixed to small but positive values, the values of the parameters $M$ and $\alpha_{s}$ obtained from the constrained fit get closer to those found for $M_{\tx{lat}} = 36-90$~MeV (Tab.~III), further evidence that $M_{B}>0$ is a more consistent choice when compared to the raw result of the fit.
\begin{table}[t]
\setlength{\tabcolsep}{10pt}
\def\arraystretch{1.5}
\begin{tabular}{c||c}
$M_{\tx{lat}}$&$p_{0}$\\
\hline
\hline
$18$&$\pm404.9 \pm 187.5 i$\\
$18^{\star}$&$\pm373.7 \pm 202.3 i$\\
$36$&$\pm388.0 \pm 194.2 i$\\
$54$&$\pm390.7 \pm 185.6 i$\\
$72$&$\pm407.7 \pm 174.9 i$\\
$90$&$\pm424.4 \pm 177.3 i$
\end{tabular}
\caption{Poles $p_{0}$ of the quark propagator derived in the minimalistic scheme, using the parameters in Tabs.~II-III. Both $M_{\tx{lat}}$ and $p_{0}$ are in MeV; the $\pm$ signs in $p_{0}$ are independent from one another. The asterisked row was obtained at fixed $M_{B}$.}
\end{table}

Being in possession of analytic expressions which give a good description of the quark mass function in the Euclidean space, we are in a position to extend the quark propagator to the complexified Minkowski space and look for its poles $p_{0}^{2}$. These are defined as the solutions to the equation
\BE
p^{2}_{0}-\mc{M}^{2}(p^{2}_{0})=0,
\EE
where the argument $p^{2}$ of the function $\mc{M}(p^{2})$ is a complexified Minkowski momentum squared, at variance with the convention used in this section, where we used the Euclidean momentum. For all the considered lattice masses, using the parameters in Tabs.~II-III, we found that the quark propagator has a pair of complex-conjugate poles in the variable $p^{2}$ (equivalently, two pairs in the variable $p=\sqrt{p^{2}}$); their positions $p_{0}$ are reported in Tab.~V. In the literature, the existence of complex-conjugate poles has been interpreted as proof of confinement, since the imaginary part of the poles has the effect of removing the particles from the asymptotic states of the theory \cite{stingl,damp,xigauge}. In the minimalistic scheme, the real part of the poles was found to be between $388$ and $424$~MeV, while their imaginary part is roughly half these values, having been found in the range from $174$ to $194$~MeV. Fixing $M_{B}=10$~MeV by hand for the lattice mass $M_{\tx{lat}}=18$~MeV yields $p_{0}=\pm373.7 \pm 202.3 i$~MeV, a result which is more consistent with those of the other lattice masses, when compared with the one obtained from the raw fit. Indeed, we note that $|\tx{Re}(p_{0})|$ increases with $M_{\tx{lat}}$, while $|\tx{Im}(p_{0})|$ decreases with it. We checked that using small but positive values of $M_{B}$ for $M_\tx{lat}=18$~MeV yields similar poles to those reported above.

\begin{figure}[t]
\centering
\includegraphics[width=0.30\textwidth,angle=-90]{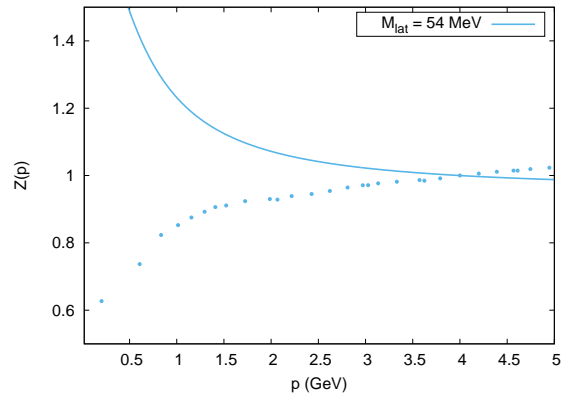}
\caption{Quark $Z$-function $Z(p^{2})$ in the Euclidean space and in the Landau gauge for $M_{\tx{lat}}=54$~MeV, renormalized at $\mu=4$~GeV. Points: quenched lattice data from Ref.~\cite{kamleh}. Curve: one-loop $Z$-function computed in the minimalistic resummation scheme using the parameters in Tab.~III.}
\end{figure}

In Fig.~9 we show an example of the $Z$-function computed in the minimalistic scheme using the parameters in Tab.~III, compared with the lattice data for a quark with mass $M_{\tx{lat}}=54$~MeV. As we can see, the behavior of $Z(p^{2})$ is the complete opposite of that found on the lattice: while on the lattice the $Z$-function increases with momentum, in the minimalistic scheme it decreases. This behavior is independent of the considered lattice mass, and we checked that it does not change if the parameters are fixed by fitting the $Z$-function itself rather than the mass function. We believe that the mismatch with the lattice data may be due to the fact that -- at least at sufficiently high energies -- $Z(p^{2})\approx 1$, making the $Z$-function very sensitive to higher-order and even non-perturbative corrections. This is supported by the results we obtained in the complex-conjugate resummation scheme, which show an improved agreement at large momenta (see Sec.~IVC ahead), and by recent findings reported in Ref.~\cite{barrios21}, where the $Z$-function is computed in the context of the Curci-Ferrari model and shown to change its behavior at two loops.\\
\begin{figure}[t]
\centering
\includegraphics[width=0.30\textwidth,angle=-90]{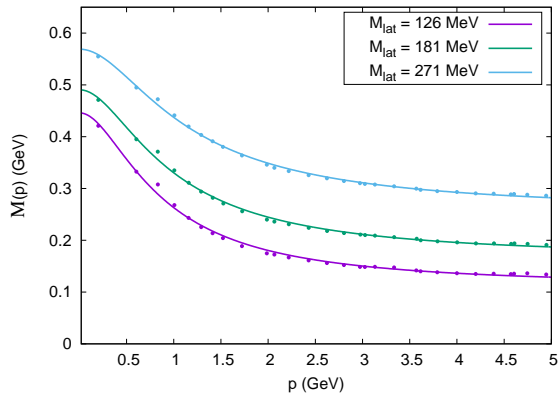}
\caption{Quark mass function $\mc{M}(p^{2})$ in the Euclidean space and in the Landau gauge for larger lattice masses $M_{\tx{lat}}$. Points: quenched lattice data from Ref.~\cite{kamleh}. Curves: one-loop mass functions computed in the minimalistic resummation scheme. The chiral masses $M$ are in the range $366-518$~MeV, while the bare masses $M_{B}$ are in the range $147-301$~MeV.}
\end{figure}

While up to this point our main focus has been on the light quarks, it may be interesting to see what happens if we try to apply the screened expansion to heavier quarks. Therefore, to end this section, we compare the minimalistic scheme mass function with the lattice data for quarks of mass $M_{\tx{lat}}=126,181,271$~MeV. The outcome is shown in Fig.~10; as in Fig.~7, the free parameters are fitted from the data themselves. It should be noted that when $M_{B}$ becomes of the same order as $M$, as is the case in these fits, the approximation that we employed throughout this paper -- namely, to neglect the finite part of diagram (2c) in Fig.~4 -- becomes less justifiable, and the diagram should be fully included in the quark self-energy. Nevertheless, it appears that the mass functions in the minimalistic scheme still manage to fit well the lattice data. As for the light quarks, the $Z$-functions computed in the minimalistic scheme for the heavier quark do not match the lattice data, and are thus not reported.

\subsection{Vertex-wise scheme}

In the vertex-wise resummation scheme, the loop diagrams included in the quark self-energy are those denoted by (2a), (2b), (2d) and, for defining a renormalized mass $M_{R}$, (2c) in Fig.~4. The quark mass function $\mc{M}(p^{2})$ can be expressed as
\begin{align}\label{massk0h0vw}
\mc{M}(p^{2})=\frac{3[k_{0}+\sigma_{S}^{(\tx{v.})}(p^{2})]}{h_{0}-\sigma_{V}^{(\tx{v.})}(p^{2})},
\end{align}
where the analytic expressions for the scalar functions $\sigma_{S}^{(\tx{v.})}(p^{2})$ and $\sigma_{V}^{(\tx{v.})}(p^{2})$ are reported in Appendix~A. As in Sec.~IVA, we fixed $m=655.7$~MeV and performed a fit to the quenched lattice mass functions of Ref.~\cite{kamleh} for the lattice masses $M_{\tx{lat}} = 18,36,54,72,90$~MeV. The results of the fit are reported in Tab.~VI, while in Tab.~VII we list the corresponding values of $\alpha_{s}$, $M_{B}$ and $M_{R}$.

No significant change was found in the behavior of the mass and $Z$-functions computed in the vertex-wise scheme when compared to the minimalistic scheme, the main difference between the two being the fitted values of the free parameters. For this reason, in what follows we will keep the discussion to a minimum and limit ourselves to reporting our results. We refer to Sec.~IVA for details.

\begin{table}[t]
\setlength{\tabcolsep}{10pt}
\def\arraystretch{1.5}
\begin{tabular}{c||c|c|c}
$M_{\tx{lat}}$&$M$&$h_{0}$&$k_{0}$\\
\hline
\hline
$18$&$268.0$&$2.656$&$-16.9$\\
$18^{\star}$&$197.6$&$2.051$&$6.8$\\
$36$&$228.7$&$2.418$&$11.5$\\
$54$&$221.4$&$2.577$&$40.0$\\
$72$&$238.4$&$2.977$&$70.1$\\
$90$&$249.0$&$3.207$&$102.5$
\end{tabular}
\caption{Fit parameters for the quark mass function $\mc{M}(p^{2})$ in the vertex-wise scheme. $M_{\tx{lat}}$, $M$ and $k_{0}$ are expressed in MeV. The lattice data are taken from Ref.~\cite{kamleh}. The asterisked row was obtained at fixed $M_{B}$, see Tab.~VII.}
\end{table}
\begin{table}[t]
\setlength{\tabcolsep}{10pt}
\def\arraystretch{1.5}
\begin{tabular}{c||c|c|c|c}
$M_{\tx{lat}}$&$M$&$\alpha_{s}$&$M_{B}$&$M_{R}$\\
\hline
\hline
$18$&$268.0$&$2.605$&$-19.1$&$-14.0$\\
$18^{\star}$&$197.6$&$3.128$&$10$&$6.8$\\
$36$&$228.7$&$2.788$&$14.3$&$10.2$\\
$54$&$221.4$&$2.663$&$46.6$&$33.9$\\
$72$&$238.4$&$2.393$&$70.7$&$53.4$\\
$90$&$249.0$&$2.261$&$95.9$&$73.8$
\end{tabular}
\caption{Fit parameters for the quark mass function $\mc{M}(p^{2})$ in the vertex-wise scheme, in terms of $\alpha_{s}$ and $M_{B}$ or $M_{R}$ (renormalization scale: $\mu=4$~GeV). $M_{\tx{lat}}$, $M$, $M_{B}$ and $M_{R}$ are expressed in MeV. The lattice data are taken from Ref.~\cite{kamleh}. The asterisked row was obtained at fixed $M_{B}$.}
\end{table}

\begin{figure}[t]
\centering
\includegraphics[width=0.30\textwidth,angle=-90]{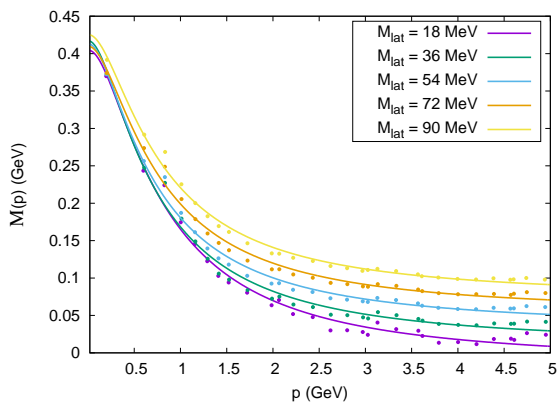}
\caption{Quark mass function $\mc{M}(p^{2})$ in the Euclidean space and in the Landau gauge for different values of the lattice mass $M_{\tx{lat}}$. Points: quenched lattice data from Ref.~\cite{kamleh}. Curves: one-loop mass functions computed in the vertex-wise resummation scheme using the parameters in Tab.~VI (equivalently, Tab.~VII).}
\end{figure}

In Fig.~11 we show the mass function $\mc{M}(p^{2})$ computed in the vertex-wise scheme together with the lattice data. As we can see, the mass functions have the same behavior as in the minimalistic scheme, and fit very well the data. Like in the former scheme, the fitted values of the bare masses $M_{B}$ are close to $M_{\tx{lat}}$, as expected upon inspection of the high-momentum limit $p^{2}\gg m^{2},M^{2}$, which in the case of the vertex-wise scheme reads
\begin{align}
\nn\sigma_{V}^{\tx{(v.)}}(p^{2})&\to-1+\frac{3m^{2}}{2p^{2}}\to-1,\\
\sigma_{S}^{\tx{(v.)}}(p^{2})&\to\frac{m^{2}}{p^{2}}\ln\frac{p^{2}}{m^{2}}+\frac{2M^{2}}{p^{2}}\ln\frac{p^{2}}{M^{2}}\to0,
\end{align}
again yielding
\BE
\mc{M}(p^{2})\to\frac{M_{B}Z_{\psi}}{Z_{\psi}+\frac{\alpha_{s}}{3\pi}}\approx M_{B}\qquad (p^{2}\gg m^{2},M^{2}).
\EE
In the vertex-wise scheme, the fitted values of the chiral mass $M$ turn out to be smaller than those reported in Sec.~IVA, being found in the range $221-249$~MeV. Together with the values of the coupling constant $\alpha_{s}$, which are larger in the minimalistic scheme, this is by far the biggest difference between the two schemes.

Like in the minimalistic scheme, the bare mass $M_{B}$ fitted from the lattice dataset $M_{\tx{lat}}=18$~MeV is negative. Again, as shown in Fig.~12, small but positive values of $M_{B}$ yield a mass function which fits well the lattice data and whose parameters $M$, $\alpha_{s}$ and $M_{R}$ are closer to those extracted from the other fits (Tab.~VII).

\begin{table}[t]
\setlength{\tabcolsep}{10pt}
\def\arraystretch{1.5}
\begin{tabular}{c|c||c|c|c}
$M_{\tx{lat}}$&$M_{B}$&$M$&$\alpha_{s}$&$M_{R}$\\
\hline
\hline
$18$&$0$&$220.9$&$2.931$&$0.0$\\
$18$&$10$&$197.6$&$3.128$&$6.8$\\
$18$&$18$&$179.7$&$3.300$&$11.9$
\end{tabular}
\caption{Fit parameters for the quark mass function $\mc{M}(p^{2})$ in the vertex-wise scheme, in terms of $\alpha_{s}$ and $M_{B}$ or $M_{R}$ (renormalization scale: $\mu=4$~GeV), given $M_{\tx{lat}} = 18$~MeV and $M_{B}$ fixed to three different values. $M_{\tx{lat}}$, $M_{B}$, $M$ and $M_{R}$ are expressed in MeV. The lattice data are taken from Ref.~\cite{kamleh}.}
\end{table}
\begin{figure}[t]
\centering
\includegraphics[width=0.30\textwidth,angle=-90]{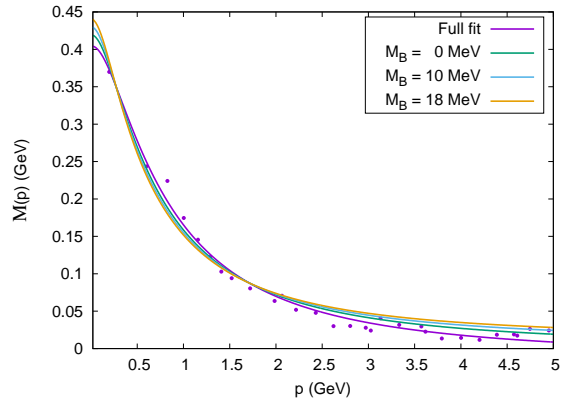}
\caption{$M_{\tx{lat}}=18$~MeV quark mass function in the Euclidean space and in the Landau gauge. Points: quenched lattice data from Ref.~\cite{kamleh}. Curves: one-loop mass functions computed in the vertex-wise resummation scheme. The parameters for the curves with $M_{B}=0,10,18$~MeV are reported in Tab.~VIII; those for the curve labeled as ``full fit'' are reported in Tab.~VII.}
\end{figure}

In Tab.~IX we report the position of the poles of the vertex-wise scheme quark propagator, obtained by using the parameters in Tab.~VII. These have real parts in the range from $371$ to $410$~MeV and imaginary parts between $167$ and $185$~MeV, slightly less than their minimalistic scheme analogues. At variance with the minimalistic scheme, we found that $|\tx{Im}(p_{0})|$ is smaller for $M_{\tx{lat}}=72$~MeV than for $M_{\tx{lat}}=90$~MeV, the difference being of few MeVs. Given the generally decreasing behavior of $|\tx{Im}(p_{0})|$ with $M_{\tx{lat}}$, we believe that this result maybe a glitch of the fit. Indeed, we checked that slightly changing the values of the free parameters for either of the two quark masses yields both a decreasing $|\tx{Im}(p_{0})|$ and mass functions which still fit well the lattice data. As for the $M_{\tx{lat}}=18$~MeV quark, if we fix $M_{B}$ to $10$~MeV like we did in Sec.~IVA, the poles are found at $p_{0}=\pm349.2 \pm 193.1i$~MeV. Again, this result is consistent with the increasing (resp. decreasing) behavior of $|\tx{Re}(p_{0})|$ (resp. $|\tx{Im}(p_{0})|$) with $M_{\tx{lat}}$, and choosing other small but positive values for $M_{B}$ does not change the picture.
\begin{table}[t]
\setlength{\tabcolsep}{10pt}
\def\arraystretch{1.5}
\begin{tabular}{c||c}
$M_{\tx{lat}}$&$p_{0}$\\
\hline
\hline
$18$&$\pm387.4 \pm 180.9 i$\\
$18^{\star}$&$\pm349.2 \pm 193.1i$\\
$36$&$\pm371.7 \pm 185.4 i$\\
$54$&$\pm375.2 \pm 177.2 i$\\
$72$&$\pm392.9 \pm 167.6 i$\\
$90$&$\pm410.8 \pm 170.2 i$
\end{tabular}
\caption{Poles $p_{0}$ of the quark propagator derived in the vertex-wise scheme, using the parameters in Tabs.~VI-VII. Both $M_{\tx{lat}}$ and $p_{0}$ are in MeV; the $\pm$ signs in $p_{0}$ are independent from one another. The asterisked row was obtained at fixed $M_{B}$.}
\end{table}
\begin{figure}[t]
\centering
\includegraphics[width=0.30\textwidth,angle=-90]{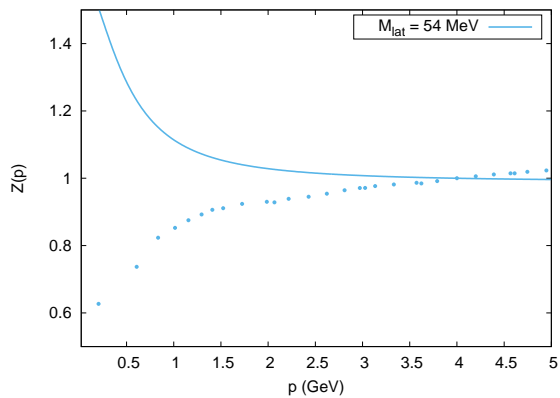}
\caption{Quark $Z$-function $Z(p^{2})$ in the Euclidean space and in the Landau gauge for $M_{\tx{lat}}=54$~MeV, renormalized at $\mu=4$~GeV. Points: quenched lattice data from Ref.~\cite{kamleh}. Curve: one-loop $Z$-function computed in the vertex-wise resummation scheme using the parameters in Tab.~VII.}
\end{figure}

The $Z$-function computed in the vertex-wise scheme, displayed in Fig.~13 for the lattice mass $M_{\tx{lat}}=54$~MeV, shows the same behavior as its minimalistic scheme counterpart, being a decreasing function of momentum. In particular, the change of scheme does not manage to solve the mismatch with the lattice data.

Finally, as in Sec.~IVA, the mass functions obtained from a fit of the heavier quarks -- $M_{\tx{lat}}=126,181,271$~MeV, see Fig.~14 -- are in good agreement with the lattice data, despite having neglected the finite part of diagram (2c) in Fig.~4.\\

We conclude that, when used to compute the quark propagator in the Landau gauge, the minimalistic and vertex-wise resummation schemes are practically equivalent: albeit with different values of the free parameters, they both yield mass functions which are found to be in good agreement with the lattice, while not being able to reproduce the correct behavior of the lattice $Z$-function. As we will see in the following section, the complex-conjugate scheme offers a partial solution to the latter issue.
\begin{figure}[t]
\centering
\includegraphics[width=0.30\textwidth,angle=-90]{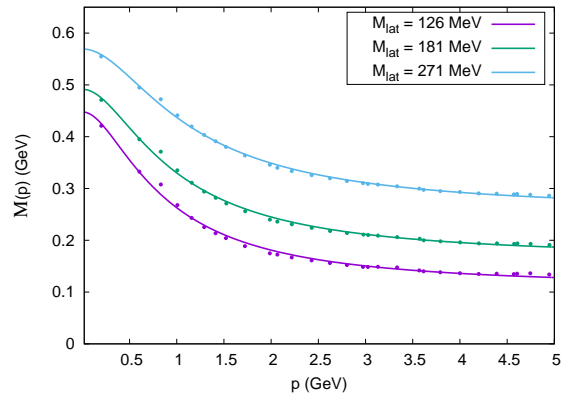}
\caption{Quark mass function $\mc{M}(p^{2})$ in the Euclidean space and in the Landau gauge for larger lattice masses $M_{\tx{lat}}$. Points: quenched lattice data from Ref.~\cite{kamleh}. Curves: one-loop mass functions computed in the vertex-wise resummation scheme. The chiral masses $M$ are in the range $288-472$~MeV, while the bare masses $M_{B}$ are in the range $136-290$~MeV.}
\end{figure}

\subsection{CC scheme}

Before reporting the results of the fits in the complex-conjugate resummation scheme, let us address one final aspect of its definition. Recall that in the CC scheme the free gluon propagator (internal gluon line) $\Delta_{\mu\nu}^{(\tx{c.c.})}(p)$ is defined modulo the absolute value of the residue $R$ of the corresponding dressed propagator at its poles. As discussed in Sec.~IIIB, since to one loop $|R|$ is multiplied to the coupling constant $\alpha_{s}$, a change in the former can be always compensated by a change in the latter. Therefore, fixing the value of $|R|$ actually amounts to choosing a definition for the coupling. In order to choose our conventions for $R$ and $\alpha_{s}$, let us inspect the divergences of the CC scheme. From Eq.~\eqref{ccms} we know that, to one loop and in the Landau gauge, the only divergence that arises in the CC scheme comes from the scalar part of the quark self-energy, and in particular from diagram (2c) in Fig.~6. Using Eq.~\eqref{SS2c}, it is easy to show that in the presence of diagram (2c)
\BE
\Sigma_{S}^{\tx{(c.c.)}}(p^{2})=\frac{\alpha_{s}}{\pi}\ \left[\sigma_{S}^{(\tx{c.c.})}(p^{2})+M_{B}Z_{\psi}\,(R+\overline{R})\,\frac{2}{\epsilon}\right],
\EE
where $\Sigma_{S}^{\tx{(c.c.)}}(p^{2})$ is the scalar part of the loop self-energy in the CC scheme and
\BE\label{sigmascc}
\sigma_{S}^{(\tx{c.c.})}(p^{2})=R\,\sigma_{S}^{(\tx{m.})}(p^{2})\Big|_{m^{2}=p_{0}^{2}}+\overline{R}\,\sigma_{S}^{(\tx{m.})}(p^{2})\Big|_{m^{2}=\overline{p_{0}^{2}}},
\EE
$\sigma_{S}^{(\tx{m.})}(p^{2})$ being the minimalistic scheme scalar function defined in Sec.~IVA. As we can see, for general values of $R=|R|e^{i\theta}$, the divergence in $\Sigma_{S}^{\tx{(c.c.)}}(p^{2})$ is not the standard one-loop divergence of QCD: a factor of $(R+\overline{R})=2|R|\cos\theta$ appears in front of the ordinary result. This is not an inconsistency by itself. As explained in Sec.~IIIB, the CC scheme is to be interpreted as a resummation of higher-order gluon polarization diagrams, so that the structure of its divergent part does not need to coincide with what we would expect from one-loop standard perturbation theory. Nonetheless, we can exploit the freedom in the choice of $|R|$ to make the scalar divergence look like a standard one-loop divergence. This can be achieved by setting
\BE
R+\overline{R}=2|R|\cos\theta = 1.
\EE
With $R$ normalized as such, we have that
\BE
\Delta_{\mu\nu}^{(\tx{c.c.})}(p)\to\frac{-it_{\mu\nu}(p)}{p^{2}}
\EE
in the UV ($p^{2}\gg m^{2}$), as in standard perturbation theory. We remark that this choice is not dictated by any profound principle that needs to be satisfied in order for the scheme to be consistent. It must be interpreted as a convention by which we fix the value of the strong coupling constant $\alpha_{s}$.\\

Having fully defined the CC scheme, let us now turn to the results of the fit. As in Secs.~IVA and IVB, the quark mass function $\mc{M}(p^{2})$ computed in the complex-conjugate scheme can be expressed as
\begin{align}\label{massk0h0cc}
\mc{M}(p^{2})=\frac{3[k_{0}+\sigma_{S}^{(\tx{c.c.})}(p^{2})]}{h_{0}-\sigma_{V}^{(\tx{c.c.})}(p^{2})},
\end{align}
where $\sigma_{S}^{(\tx{c.c.})}(p^{2})$ is given by Eq.~\eqref{sigmascc} and
\BE\label{sigmavcc}
\sigma_{V}^{(\tx{c.c.})}(p^{2})=R\,\sigma_{V}^{(\tx{m.})}(p^{2})\Big|_{m^{2}=p_{0}^{2}}+\overline{R}\,\sigma_{V}^{(\tx{m.})}(p^{2})\Big|_{m^{2}=\overline{p_{0}^{2}}},
\EE
$\sigma_{V}^{(\tx{m.})}(p^{2})$ having been defined in Sec.~IVA. In order to fix the value of the free parameters $k_{0}$ and $h_{0}$, we fitted the quenched lattice mass functions of Ref.~\cite{kamleh} for the quark masses $M_{\tx{lat}} = 18,36,54,72,90$~MeV, using $m=655.7$~MeV as the gluon mass parameter. The results of the fit are reported in Tabs.~X and XI.\\

\begin{table}[t]
\setlength{\tabcolsep}{10pt}
\def\arraystretch{1.5}
\begin{tabular}{c||c|c|c}
$M_{\tx{lat}}$&$M$&$h_{0}$&$k_{0}$\\
\hline
\hline
$18$&$449.9$&$6.294$&$-4.6$\\
$18^{\star}$&$405.9$&$5.467$&$18.2$\\
$36$&$406.6$&$5.701$&$49.0$\\
$54$&$405.2$&$6.166$&$108.0$\\
$72$&$431.9$&$7.216$&$176.3$\\
$90$&$449.8$&$7.801$&$248.3$
\end{tabular}
\caption{Fit parameters for the quark mass function $\mc{M}(p^{2})$ in the complex-conjugate scheme. $M_{\tx{lat}}$, $M$ and $k_{0}$ are expressed in MeV. The lattice data are taken from Ref.~\cite{kamleh}. The asterisked row was obtained at fixed $M_{B}$, see Tab.~XI.}
\end{table}
\begin{table}[t]
\setlength{\tabcolsep}{10pt}
\def\arraystretch{1.5}
\begin{tabular}{c||c|c|c|c}
$M_{\tx{lat}}$&$M$&$\alpha_{s}$&$M_{B}$&$M_{R}$\\
\hline
\hline
$18$&$449.9$&$1.252$&$-2.2$&$-1.8$\\
$18^{\star}$&$405.9$&$1.407$&$10$&$8.2$\\
$36$&$406.6$&$1.359$&$25.8$&$21.2$\\
$54$&$405.2$&$1.273$&$52.6$&$43.8$\\
$72$&$431.9$&$1.115$&$73.3$&$62.6$\\
$90$&$449.8$&$1.043$&$95.5$&$82.4$
\end{tabular}
\caption{Fit parameters for the quark mass function $\mc{M}(p^{2})$ in the complex-conjugate scheme, in terms of $\alpha_{s}$ and $M_{B}$ or $M_{R}$ (renormalization scale: $\mu=4$~GeV). $M_{\tx{lat}}$, $M$, $M_{B}$ and $M_{R}$ are expressed in MeV. The lattice data are taken from Ref.~\cite{kamleh}. The asterisked row was obtained at fixed $M_{B}$.}
\end{table}
\begin{figure}[t]
\centering
\includegraphics[width=0.30\textwidth,angle=-90]{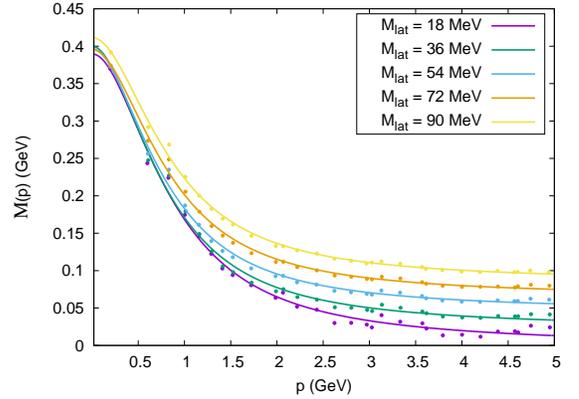}
\caption{Quark mass function $\mc{M}(p^{2})$ in the Euclidean space and in the Landau gauge for different values of the lattice mass $M_{\tx{lat}}$. Points: quenched lattice data from Ref.~\cite{kamleh}. Curves: one-loop mass functions computed in the complex-conjugate resummation scheme using the parameters in Tab.~X (equivalently, Tab.~XI).}
\end{figure}

In Fig.~15 we show the complex-conjugate scheme mass functions $\mc{M}(p^{2})$ together with the lattice data. As in the minimalistic and vertex-wise schemes, our analytic functions are in very good agreement with the data. The chiral mass $M$ is found in the range from $405$ to $450$~MeV, and the values of $M_{B}$ increase with $M_{\tx{lat}}$: having set $2|R|\cos\theta=1$ makes Eqs.~\eqref{UVA} and \eqref{UVB} hold also in the CC scheme. For the $M_{\tx{lat}}=18$~MeV quark, which by a raw fit, as in the previous schemes, is found to have negative bare mass, fixing $M_{B}$ to small but positive values still results in a mass function which fits well the lattice data -- see Tab.~XII and Fig.~16.
\begin{table}[t]
\setlength{\tabcolsep}{10pt}
\def\arraystretch{1.5}
\begin{tabular}{c|c||c|c|c}
$M_{\tx{lat}}$&$M_{B}$&$M$&$\alpha_{s}$&$M_{R}$\\
\hline
\hline
$18$&$0$&$441.6$&$1.279$&$0.0$\\
$18$&$10$&$405.9$&$1.407$&$8.2$\\
$18$&$18$&$379.5$&$1.519$&$14.4$
\end{tabular}
\caption{Fit parameters for the quark mass function $\mc{M}(p^{2})$ in the complex-conjugate scheme, in terms of $\alpha_{s}$ and $M_{B}$ or $M_{R}$ (renormalization scale: $\mu=4$~GeV), given $M_{\tx{lat}} = 18$~MeV and $M_{B}$ fixed to three different values. $M_{\tx{lat}}$, $M_{B}$, $M$ and $M_{R}$ are expressed in MeV. The lattice data are taken from Ref.~\cite{kamleh}.}
\end{table}
\begin{figure}[t]
\centering
\includegraphics[width=0.30\textwidth,angle=-90]{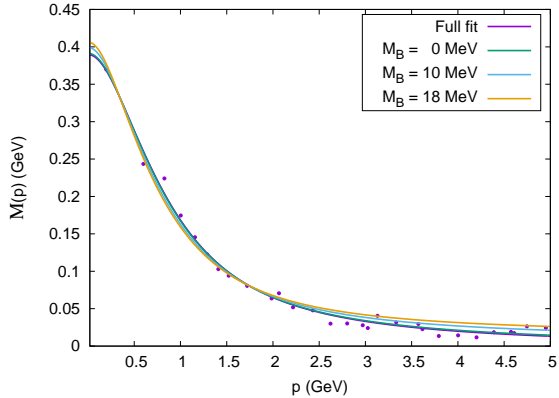}
\caption{$M_{\tx{lat}}=18$~MeV quark mass function in the Euclidean space and in the Landau gauge. Points: quenched lattice data from Ref.~\cite{kamleh}. Curves: one-loop mass functions computed in the complex-conjugate resummation scheme. The parameters for the curves with $M_{B}=0,10,18$~MeV are reported in Tab.~XII; those for the curve labeled as ``full fit'' are reported in Tab.~XI.}
\end{figure}

The CC quark propagator has a pair of complex-conjugate poles, whose positions are reported in Tab.~XIII. With $M_{B}$ fixed to example value of $10$~MeV, $|\tx{Re}(p_{0})|$ is found in the range from $423$ to $478$~MeV, increasing with $M_{\tx{lat}}$, while $|\tx{Im}(p_{0})|$ lies between $186$ and $157$~MeV, decreasing with it. The former are quite larger than those of the minimalistic and vertex-wise schemes, while the latter are somewhat smaller. In other words, the ratio $|\tx{Im}(p_{0})/\tx{Re}(p_{0})|$ tends to be smaller in the CC scheme in comparison to the other schemes.
\begin{table}[t]
\setlength{\tabcolsep}{10pt}
\def\arraystretch{1.5}
\begin{tabular}{c||c}
$M_{\tx{lat}}$&$p_{0}$\\
\hline
\hline
$18$&$\pm448.8\pm167.9i$\\
$18^{\star}$&$\pm423.8 \pm 186.0i$\\
$36$&$\pm428.5\pm182.4i$\\
$54$&$\pm434.2\pm172.5i$\\
$72$&$\pm457.1\pm155.7i$\\
$90$&$\pm477.7\pm157.6i$
\end{tabular}
\caption{Poles $p_{0}$ of the quark propagator derived in the complex-conjugate scheme, using the parameters in Tabs.~X-XI. Both $M_{\tx{lat}}$ and $p_{0}$ are in MeV; the $\pm$ signs in $p_{0}$ are independent from one another. The asterisked row was obtained at fixed $M_{B}$.}
\end{table}

Along with some differences in the fitted values of the free parameters and in the position of the quark poles, the mass functions computed in the CC scheme also show a small change in shape, when compared to their analogues in the minimalistic and vertex-wise schemes. This is displayed in Fig.~17, where we plot the mass functions obtained in the three schemes for the example value of $M_{\tx{lat}}=54$~MeV. As a result of the change, the CC scheme mass function is somewhat more suppressed in the $p\to 0$ limit. The effect, however, is very small and might not be meaningful.

\begin{figure}[t]
\centering
\includegraphics[width=0.30\textwidth,angle=-90]{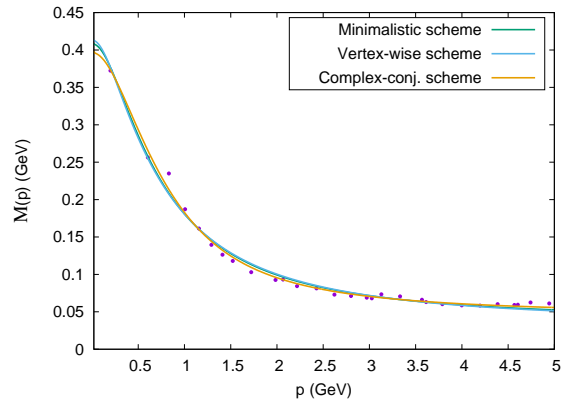}
\caption{Quark mass function $\mc{M}(p^{2})$ in the Euclidean space and in the Landau gauge for $M_{\tx{lat}}=54$~MeV. Points: quenched lattice data from Ref.~\cite{kamleh}. Curves: one-loop mass functions computed in the minimalistic, vertex-wise and complex-conjugate resummation schemes.}
\end{figure}

The radical departure of the complex-conjugate scheme from the minimalistic and vertex-wise schemes concerns the $Z$-function. In Fig.~18 we plot $Z(p^{2})$ for the example value of $M_{\tx{lat}}=54$~MeV together with the lattice data. As we can see, at variance with the previous two schemes and consistent with the lattice, the CC scheme $Z$-function increases with momentum for $p\gtrapprox1$~GeV. Moreover, above this cutoff value, our analytical expression is also in fair quantitative agreement with the lattice data\footnote{Observe that in Fig.~18 the $Z$-function is plotted on an enlarged scale: for $p>1.0-1.5$~GeV the difference between the function computed in the CC scheme and The lattice data are at most around $10-20\%$.}. At low momenta, on the other hand, the agreement is lost, since $Z(p^{2})$ changes behavior and starts to increase with decreasing $p$. This picture holds for any of the lattice masses considered in this section.

It appears that, at sufficiently large momenta, computing the quark $Z$-function with the fully dressed gluon propagator (or, to be more precise, its CC scheme approximation) as the internal gluon line of the quark self-energy solves the mismatch between the screened expansion and the lattice data. As discussed in Sec.~IIIB, this may be due to the dressed gluon propagator containing non-perturbative contributions (e.g. from the condensates, consistent with the OPE studies \cite{arriola,blossier,wang}) which a bare massive propagator does not.\\
\begin{figure}[t]
\centering
\includegraphics[width=0.30\textwidth,angle=-90]{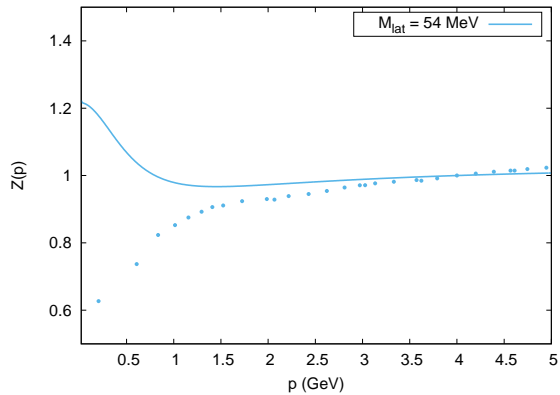}
\caption{Quark $Z$-function $Z(p^{2})$ in the Euclidean space and in the Landau gauge for $M_{\tx{lat}}=54$~MeV, renormalized at $\mu=4$~GeV. Points: quenched lattice data from Ref.~\cite{kamleh}. Curve: one-loop $Z$-function computed in the complex-conjugate resummation scheme using the parameters in Tab.~XI.}
\end{figure}

To end this section, as we did in Sec.~IVA and IVB, in Fig.~19 we compare the mass function with the lattice data for heavier quarks, $M_{\tx{lat}}=126,181,271$~MeV. We see that also in the CC scheme our analytic expressions fit well the data.\\
\begin{figure}[t]
\centering
\includegraphics[width=0.30\textwidth,angle=-90]{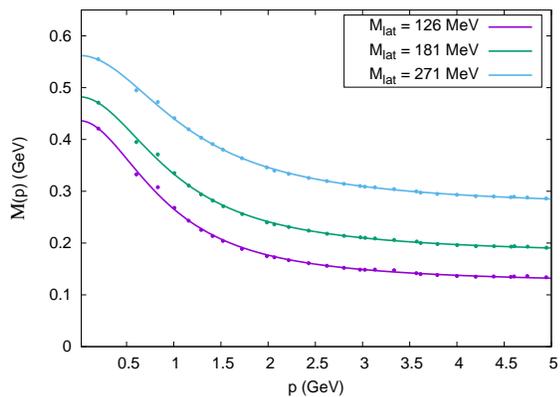}
\caption{Quark mass function $\mc{M}(p^{2})$ in the Euclidean space and in the Landau gauge for larger lattice masses $M_{\tx{lat}}$. Points: quenched lattice data from Ref.~\cite{kamleh}. Curves: one-loop mass functions computed in the complex-conjugate resummation scheme. The chiral masses $M$ are in the range $503-738$~MeV, while the bare masses $M_{B}$ are in the range $132-282$~MeV.}
\end{figure}

\section{Discussion}
 
The present work was motivated by the ambitious aim of developing a reliable analytical approach to non-perturbative QCD from first principles. In this paper, important progresses have been made by the inclusion of quarks in the successful framework of the screened expansion, which was first introduced for pure YM theory in~\cite{ptqcd,ptqcd2}. Here we have shown that, without any change to the gauge-fixed Faddeev-Popov Lagrangian, by a wise choice of the expansion point and by a reasonable setting of the scheme and parameters, perturbation theory gives a quantitative agreement with the available lattice data for the quark mass function -- albeit in the quenched case until now. This constitutes an improvement over the results of a previous analysis, which led to an only qualitative description of the quark sector \cite{analyt}.

Because of the agreement which is reached with the lattice in the Euclidean space, we believe that the analytic properties of the mass function might be reliable in the whole complex plane up to moderately high energies. Thus, the explicit one-loop analytical expressions are not just good interpolation formulas, but they also unveil important analytic features of the propagators, like the existence of complex-conjugate poles, pointing to a confinement scenario which is rooted in those peculiar features which make quarks and gluons unobservable, yielding a dynamical mechanism for their exclusion from the asymptotic states.

While the existence of complex-conjugated poles might not be a direct proof of confinement \cite{kondo2021}, their existence would be ruled out if quarks were present in the asymptotic states. Actually, the usual K\"{a}llen-Lehmann relations do not hold if there are complex poles and the relative spectral densities do not satisfy the usual positivity conditions.

We must note that in Ref.~\cite{analyt} -- which used the same formalism of the present paper, albeit in a different scheme, to study the chiral limit of QCD -- the quark propagator was found to have a unique pole on the real axis. In that work, as we said, the agreement with the lattice data was only qualitative: the data themselves showed large error bars and fluctuations, so that any comparison with the analytic result could not be conclusive. Having attained a much better match with the lattice now leads us to revisit our previous results.\\

Unfortunately, our main aim is far from being fully achieved yet, and, despite the good quantitative description of the quark mass function, many aspects must still be addressed. First of all, we must still find a way to fix from first principles the two spurious parameters which arise from the approximation, namely an arbitrary additive constant which emerges from the renormalization of the one-loop quark self-energy and the ratio $M/m$ between the quark and the gluon mass scales, which are arbitrary up to an overall choice for the energy units. 

In pure YM theory, by enforcing some constraints of BRST symmetry, like the Nielsen identities \cite{nielsen, kobes90,breck}, the expansion can be optimized yielding a fully predictive method which does not require any external input and does not contain any spurious parameter \cite{xigauge}. In the quark sector, we still have to fix the spurious parameters by a fit of the available lattice data. While it is encouraging to see that an optimal choice of the parameters does exist which describes the quark mass function data very well for any given lattice mass, we still expect that the spurious parameters might be fixed by enforcing some constraints from first principles, like we did for pure YM theory. 

Of course, if carried out by employing the Nielsen identities or similar exact methods, this program would require a fully consistent calculation for the interacting quark-gluon theory. In the present approach, we instead used the optimized parameters of pure YM theory and investigated the quark sector in a quenched approximation. Even at one loop, the existence of quarks modifies the gluon polarization by a quark loop which was not included in the gluon optimization. Thus, we expect that the removal of all spurious parameters by first principles like in \cite{xigauge} will require a fully consistent, unquenched calculation.

Another important issue is the truncation of the expansion, which, in the absence of a unique smallness parameter, like the coupling in ordinary perturbation theory, might appear quite arbitrary. In principle, the method allows us to carry out the calculations perturbatively, by adding higher-order corrections; however, in order to do so, a general criterion for the order-by-order truncation of the expansion is required. In this work, we have shown that the ambiguity can only arise for finite graphs, since the cancellation of spurious divergences requires a well defined set of graphs to be retained at each order. Moreover, at one loop, the residual ambiguity seems to be compensated by a change in the values of the spurious free parameters, with basically no residual effect on the quark propagator. Even in the complex plane, the pole position is quite robust, with only a few percent change when going from a truncation scheme to the other. In this respect, the weak dependence of the pole position on the resummation scheme can be regarded as an estimate of the accuracy of the method.

Despite the difficulties, the available data for light quarks remain the most important benchmark for our predictions, since the non-perturbative effects, like dynamical mass generation and chiral symmetry breaking, become less evident for heavier quarks. Nonetheless, we checked that the agreement with the data is very good even for lattice masses in the range $100-300$~MeV.\\

A non-perturbative feature which is not captured by either the minimalistic or the vertex-wise scheme is the slightly increasing tail of the $Z$-function shown by the lattice data. This behavior can be understood by the OPE, which predicts a powerlike behavior for $Z(p^{2})$, with a coefficient proportional to the dimension-2 gluon condensate $\avg{A^{2}}$ \cite{boucaud4}. It is a pure non-perturbative effect which the present one-loop expansion fails to predict, unless some kind of resummation is performed; the same mismatch has been observed in other massive models, like the Curci-Ferrari model \cite{tissier14}. We note that, in the tail, the effects of the interactions on the lattice $Z$-function are very small, so that $Z(p^{2})\approx 1$. Thus, the observed deviations are not very relevant for the overall description of the quark propagator, which at moderately high energies is basically determined by the mass function alone. Actually, the one-loop contribution to $Z(p^{2})$, too, is finite and very small, explaining why the $Z$-function is so sensitive to higher-order corrections \cite{barrios21} and thermal effects \cite{olive19}. In the context of the Curci-Ferrari model \cite{barrios21}, it has been shown that the two-loop self-energy is enough to correct the behavior of the $Z$-function over the whole momentum range.

On the other hand, the almost vanishing perturbative contributions make $Z(p^{2})$ a very interesting benchmark for investigating non-perturbative effects and the role of the gluon condensate through the OPE at large energies. It is remarkable that, if the gluon line is resummed inside the one-loop quark self energy, replacing the free-gluon propagator with the dressed one-loop gluon line, an increasing $Z$-function is found at large momenta, just where the OPE result should hold. Since the main feature of the non-perturbative resummation is the existence of complex-conjugated poles in the dressed gluon propagator, instead of the real pole of the undressed propagator, we argue that the complex gluon poles might be related with the existence of a non-vanishing gluon condensate \cite{cucch12}.

Overall, we can say that, when optimized, the screened massive expansion provides a quantitative and analytical tool for investigating the infrared limit of the full QCD, at least in the quenched approximation. The results are very encouraging and suggest that in a fully consistent unquenched calculation, even the residual free parameters might be fixed by the general constraints of BRST symmetry, yielding a more complete analytical description of non-perturbative QCD from first principles.

\acknowledgments

This research was supported in part by ``Piano per la Ricerca di Ateneo - Linea di intervento 2'' of the University of Catania.

\appendix

\section{QUARK SELF-ENERGY}

In this appendix we report the relevant functions for the screened expansion's quark propagator in the minimalistic and vertex-wise resummation schemes. As discussed in Sec.~IIIB, the corresponding complex-conjugate scheme functions are easily derived from the minimalistic scheme; this is proven in Appendix B.

\subsection{Diagrams (2a), (2b) and (2d)}

In Euclidean space, the self-energy contribution $\Sigma^{(2a)}(p)$ due to the uncrossed quark loop, i.e. diagram (2a) in Fig.~4, can be divided into a vector and a scalar component, $\Sigma^{(2a)}_{V}(p^{2})$ and $\Sigma^{(2a)}_{S}(p^{2})$, as
\BE
\Sigma^{(2a)}(p)=i\slashed{p}\,\Sigma^{(2a)}_{V}(p^{2})+\Sigma_{S}^{(2a)}(p^{2}).
\EE
The two components can be expressed in terms of two scalar functions $\sigma_{V}^{(2a)}(p^{2})$ and $\sigma_{S}^{(2a)}(p^{2})$ as
\begin{align}\label{sig2asig}
\nn\Sigma^{(2a)}_{V}(p^{2})&=\frac{\alpha_{s}}{3\pi}\,\sigma^{(2a)}_{V}(p^{2}),\\
\Sigma^{(2a)}_{S}(p^{2})&=\frac{\alpha_{s}}{\pi}\,M\, \left\{\frac{2}{\epsilon}-\ln\frac{M^{2}}{\overline{\mu}^{2}}+\sigma^{(2a)}_{S}(p^{2})\right\},
\end{align}
where $\epsilon=4-d$ and $\overline{\mu}$ is an arbitrary scale introduced by dimensional regularization. If we define two adimensional variables $s$ and $x$, representing the Euclidean momentum $p^{2}$ and the quark chiral mass $M$,
\BE
s=p^{2}/m^{2},\qquad\qquad x=M^{2}/m^{2},
\EE
then the functions $\sigma^{(2a)}_{V}$ and $\sigma^{(2a)}_{S}$ can be put in the form
\begin{align}\label{se2a1}
\nn\sigma_{V}^{(2a)} &= C_{R}\,\ln R + C_{x}\,\ln x + C_{xs}\,\ln\frac{x}{x + s} + C_{0},\\
\sigma_{S}^{(2a)} &= \frac{t}{s}\,\ln R - \frac{t - s - x + 1}{2 s}\,\ln x,
\end{align}
where the coefficient functions $C_{R}, C_{x}, C_{xs}$ and $C_{0}$ read
\begin{align}\label{se2a2}
\nn C_{R} &= \frac{t}{2 s^{2}}\ [(x + s)^2 + (x - s) - 2],\\
\nn C_{x} &= -\frac{1}{2}\,C_{R} + \frac{1}{4 s^2}\ [(x + s)^3 - 3 (x - s) + 2],\\
\nn C_{xs} &= -\frac{(x + s)^3}{2 s^2},\\
C_{0} &= \frac{x - 2}{2 s} - \frac{1}{2},
\end{align}
while $R$ is defined as
\BE\label{se2a3}
R=\frac{t-s+x-1}{t+s+x-1}.
\EE
In Eqs.~\eqref{se2a1} to \eqref{se2a3}, $t$ is itself a function of $s$ and $x$, defined as
\BE
t=\sqrt{(x + s)^{2}+2(s-x)+1}.
\EE
The expressions reported above agree with those computed in the one-loop Curci-Ferrari model \cite{tissier14}.

As discussed in Sec.~III, diagrams (2b) and (2d) in Fig.~4 can be computed as derivatives of diagram (2a):
\begin{align}\label{sig2b2dder}
\nn\Sigma^{(2b)}(p)&=-M\frac{\partial}{\partial M}\,\Sigma^{(2a)}(p),\\
\Sigma^{(2d)}(p)&=-m^{2}\frac{\partial}{\partial m^{2}}\,\Sigma^{(2a)}(p).
\end{align}
Once split into a vector and a scalar component,
\begin{align}
\nn\Sigma^{(2b)}(p)&=i\slashed{p}\,\Sigma^{(2b)}_{V}(p^{2})+\Sigma_{S}^{(2b)}(p^{2}),\\
\Sigma^{(2d)}(p)&=i\slashed{p}\,\Sigma^{(2d)}_{V}(p^{2})+\Sigma_{S}^{(2d)}(p^{2}),
\end{align}
$\Sigma^{(2b)}(p)$ and $\Sigma^{(2d)}(p)$ can be expressed in terms of four scalar functions, $\sigma_{V,S}^{(2b)}(p^{2})$ and $\sigma_{V,S}^{(2d)}(p^{2})$:
\begin{align}
\nn\Sigma^{(2b)}_{V}(p)&=\frac{\alpha_{s}}{3\pi}\,\sigma^{(2b)}_{V}(p^{2}),\\
\nn\Sigma^{(2b)}_{S}(p)&=\frac{\alpha_{s}}{\pi}\,M\, \left\{-\frac{2}{\epsilon}+\ln\frac{M^{2}}{\overline{\mu}^{2}}+\nn\sigma^{(2b)}_{S}(p^{2})\right\},\\
\nn\Sigma^{(2d)}_{V}(p)&=\frac{\alpha_{s}}{3\pi}\,\sigma^{(2d)}_{V}(p^{2}),\\
\Sigma^{(2d)}_{S}(p)&=\frac{\alpha_{s}}{\pi}\,M\, \sigma^{(2d)}_{S}(p^{2}).
\end{align}
Using Eqs.~\eqref{sig2b2dder} and \eqref{sig2asig}, it is easy to compute these functions as derivatives of $\sigma_{V}^{(2a)}$ and $\sigma_{S}^{(2a)}$: for diagram (2b) we have
\begin{align}\label{sig2bVS2a}
\nn\sigma_{V}^{(2b)}&=-M\frac{\partial}{\partial M}\,\sigma_{V}^{(2a)},\\
\nn\sigma_{S}^{(2b)}&=-\frac{\partial}{\partial M}\,[M\sigma_{S}^{(2a)}]+2=\\
&=-\sigma_{S}^{(2a)}-M\frac{\partial}{\partial M}\,\sigma_{S}^{(2a)}+2,
\end{align}
whereas for diagram (2d)
\begin{align}\label{sig2dVS2a}
\nn\sigma_{V}^{(2d)}=-m^{2}\frac{\partial}{\partial m^{2}}\,\sigma_{V}^{(2a)},\\
\sigma_{S}^{(2d)}=-m^{2}\frac{\partial}{\partial m^{2}}\,\sigma_{S}^{(2a)}.
\end{align}
Note that the $2$ on the right-hand side of $\sigma_{S}^{(2b)}$ comes from the derivative of $\ln M^{2}$ inside the brackets in Eq.~\eqref{sig2asig}.

In what follows, we will report the explicit self-energy functions computed in the minimalistic and vertex-wise resummation schemes.

\subsection{Self-energy in the minimalistic and vertex-wise resummation schemes}

Recall that in the minimalistic scheme we only keep the self-energy diagrams (2a) and (2b), whereas in the vertex-wise scheme we also include diagram (2d). Let us start from the first one.

In the minimalistic scheme, the loop contribution $\Sigma^{\tx{(m.)}}(p)$ to the quark self-energy is given by
\BE
\Sigma^{\tx{(m.)}}(p)=\Sigma^{(2a)}(p)+\Sigma^{(2b)}(p).
\EE
If we split $\Sigma^{\tx{(m.)}}(p)$ into a vector and a scalar component,
\BE
\Sigma^{\tx{(m.)}}(p)=i\slashed{p}\,\Sigma^{\tx{(m.)}}_{V}(p^{2})+\Sigma^{\tx{(m.)}}_{S}(p^{2}),
\EE
then $\Sigma^{\tx{(m.)}}_{V}(p^{2})$ and $\Sigma^{\tx{(m.)}}_{S}(p^{2})$ can be expressed in terms of two scalar functions $\sigma^{\tx{(m.)}}_{V}(p^{2})$ and $\sigma^{\tx{(m.)}}_{S}(p^{2})$, as
\begin{align}
\nn \Sigma^{\tx{(m.)}}_{V}(p^{2})&=\frac{\alpha_{s}}{3\pi}\,\sigma_{V}^{\tx{(m.)}}(p^{2}),\\
\Sigma^{\tx{(m.)}}_{S}(p^{2})&=\frac{\alpha_{s}}{\pi}\,M\,\sigma_{S}^{\tx{(m.)}}(p^{2}).
\end{align}
Here,
\begin{align}
\nn \sigma_{V}^{\tx{(m.)}}&=\sigma_{V}^{(2a)}+\sigma_{V}^{(2b)},\\
\sigma_{S}^{\tx{(m.)}}&=\sigma_{S}^{(2a)}+\sigma_{S}^{(2b)}.
\end{align}
Going back to Eq.~\eqref{sig2bVS2a}, the derivatives with respect to $M$ can be traded with derivatives with respect to $x=M^{2}/m^{2}$,
\BE
M\frac{\partial}{\partial M}=2x\,\frac{\partial}{\partial x};
\EE
\\
then, $\sigma^{(2b)}_{V,S}$ can be expressed as the following derivatives of $\sigma^{(2a)}_{V,S}$:
\begin{align}
\nn\sigma^{\tx{(m.)}}_{V}&=\left(1-2x\frac{\partial}{\partial x}\right)\sigma^{(2a)}_{V},\\
\sigma^{\tx{(m.)}}_{S}&=-2x\frac{\partial}{\partial x}\,\sigma^{(2a)}_{S}+2.
\end{align}
A straightforward albeit tedious calculation leads to the result
\begin{widetext}
\begin{align}
\nn\sigma_{V}^{\tx{(m.)}} &= C_{R}^{\tx{(m.)}}\,\ln R + C_{x}^{\tx{(m.)}}\,\ln x + C_{xs}^{\tx{(m.)}}\,\ln\frac{x}{x + s} + C_{0}^{\tx{(m.)}},\\
\sigma^{\tx{(m.)}}_{S}&=-\frac{2x(x+s-1)}{st}\,\ln R-\frac{x(t-x-s+1)}{st}\,\ln x,
\end{align}
where the coefficient functions $C_{R}^{\tx{(m.)}}, C_{x}^{\tx{(m.)}}, C_{xs}^{\tx{(m.)}}$ and $C_{0}^{\tx{(m.)}}$ read
\begin{align}\label{sem2}
\nn C_{R}^{\tx{(m.)}} &=\frac{1}{2s^{2}t}\,\{(s - 5 x)[(s + x)^3 + (s^2 - x^2)] - 3 (s^2 - x^2) - 4 sx - 
 5 s - x - 2\},\\
\nn C_{x}^{\tx{(m.)}} &= -\frac{1}{2}\,C_{R} ^{(\tx{m.})}+ \frac{1}{4 s^2}\ [(s-5x)(x + s)^2+3(x+s)+2],\\
\nn C_{xs}^{\tx{(m.)}} &= -\frac{(x + s)^2}{2 s^2}(s-5x),\\
C_{0}^{\tx{(m.)}} &= -\frac{5x + 2}{2 s} - \frac{1}{2}.
\end{align}
\end{widetext}

Similarly, in the vertex-wise scheme, by including diagram (2d) to obtain the loop contribution $\Sigma^{(\tx{v.})}(p)$ to the self-energy,
\BE
\Sigma^{\tx{(v.)}}(p)=\Sigma^{(2a)}(p)+\Sigma^{(2b)}(p)+\Sigma^{(2d)}(p),
\EE
we can write
\BE
\Sigma^{\tx{(v.)}}(p)=i\slashed{p}\,\Sigma^{\tx{(v.)}}_{V}(p^{2})+\Sigma^{\tx{(v.)}}_{S}(p^{2}),
\EE
and express $\Sigma^{\tx{(v.)}}_{V}(p^{2})$ and $\Sigma^{\tx{(v.)}}_{S}(p^{2})$ in terms of two scalar functions $\sigma^{\tx{(v.)}}_{V}(p^{2})$ and $\sigma^{\tx{(v.)}}_{S}(p^{2})$,
\begin{align}
\nn \Sigma^{\tx{(v.)}}_{V}(p^{2})&=\frac{\alpha_{s}}{3\pi}\,\sigma_{V}^{\tx{(v.)}}(p^{2}),\\
\Sigma^{\tx{(v.)}}_{S}(p^{2})&=\frac{\alpha_{s}}{\pi}\,M\,\sigma_{S}^{\tx{(v.)}}(p^{2}).
\end{align}
Clearly,
\begin{align}
\nn \sigma_{V}^{\tx{(v.)}}&=\sigma_{V}^{(2a)}+\sigma_{V}^{(2b)}+\sigma_{V}^{(2d)},\\
\sigma_{S}^{\tx{(v.)}}&=\sigma_{S}^{(2a)}+\sigma_{S}^{(2b)}+\sigma_{S}^{(2d)}.
\end{align}
Using the previous results for $\sigma_{V,S}^{(2b)}$, together with Eq.~ \eqref{sig2dVS2a} and
\begin{align}
m^{2}\frac{\partial}{\partial m^{2}}&=-s\frac{\partial}{\partial s}-x\frac{\partial}{\partial x}, 
\end{align}
it is easy to show that the scalar functions $\sigma^{\tx{(v.)}}_{V,S}$ can be computed as the following derivatives of $\sigma^{(2a)}_{V,S}$:
\begin{align}
\nn\sigma^{\tx{(v.)}}_{V}&=\left(1-x\frac{\partial}{\partial x}+s\frac{\partial}{\partial s}\right)\sigma^{(2a)}_{V},\\
\sigma^{\tx{(v.)}}_{S}&=\left(-x\frac{\partial}{\partial x}+s\frac{\partial}{\partial s}\right)\,\sigma^{(2a)}_{S}+2.
\end{align}
A lengthy calculation yields \cite{analyt}
\begin{widetext}
\begin{align}
\nn\sigma_{V}^{\tx{(v.)}} &= C_{R}^{\tx{(v.)}}\,\ln R + C_{x}^{\tx{(v.)}}\,\ln x + C_{xs}^{\tx{(v.)}}\,\ln\frac{x}{x + s} + C_{0}^{\tx{(v.)}},\\
\sigma^{\tx{(v.)}}_{S}&=-\frac{s(2x+1)+(2x-1)(x-1)}{st}\,\ln \frac{R}{\sqrt{x}}+\frac{1-2x}{2s}\,\ln x,
\end{align}
where the coefficient functions $C_{R}^{\tx{(v.)}}, C_{x}^{\tx{(v.)}}, C_{xs}^{\tx{(v.)}}$ and $C_{0}^{\tx{(v.)}}$ read
\begin{align}\label{sev2}
\nn C_{R}^{\tx{(v.)}} &= \frac{1}{s^{2}t}\ \{(s-2x)[(x + s)^3+(s^{2}-x^{2})] + (s-x+1)(1-3x)+2sx\},\\
\nn C_{x}^{\tx{(v.)}} &= -\frac{1}{2} \, C_{R}^{(\tx{v.})}+ \frac{1}{2 s^2}\ [(x + s)^2(s-2x)+3x-1],\\
\nn C_{xs}^{\tx{(v.)}} &= -\frac{(s-2x)(x+s)^2}{s^2},\\
C_{0}^{\tx{(v.)}} &= \frac{1-2x}{s}.
\end{align}
\end{widetext}

\section{LOOP INTEGRALS IN THE CC SCHEME}

The complex-conjugate (CC) scheme for the quenched one-loop quark propagator is defined by the internal gluon lines in Fig.~6 being set equal to the principal part of the fully dressed gluon propagator: in Euclidean space
\BE
\Delta^{(\tx{c.c.})}_{\mu\nu}(p)=\left\{\frac{R}{p^{2}+p_{0}^{2}}+\frac{\overline{R}}{p^{2}+\overline{p_{0}^{2}}}\right\}\ t_{\mu\nu}(p),
\EE
where the values of $p_{0}^{2}$, $R$ and of their complex conjugates $\overline{p_{0}^{2}}$ and $\overline{R}$ are derived in the framework of the screened expansion of pure Yang-Mills theory\footnote{ The value of $|R|$ is actually inessential in our calculation -- see Sec.~IIIB.} (see Sec.~IIIB and Tab.~I in Sec.~IIB).

The loop diagrams (2a) to (2c) in Fig.~6 can be computed by employing the usual machinery of Feynman parameter integrals and Gamma functions. In order to see this, first note that the Feynman parameter formula
\BE
\frac{1}{AB}=\int_{0}^{1}dx\ \frac{1}{[xA+(1-x)B]^{2}}
\EE
remains valid for complex $A$ and $B$. As a consequence, in Euclidean space, all the loop integrals can be expressed in terms of double integrals $\mc{I}$ of the form
\BE\label{intcc1}
\mc{I}=\int_{0}^{1}dx\int\frac{d^{d}q}{(2\pi)^{d}}\frac{(q^{2})^{n}}{(q^{2}+\Delta)^{2}},
\EE
where $n$ is equal to either $0$ or $1$. In the above equation, at variance with the standard case,
\BE
\Delta=xp_{0}^{2}+(1-x)M^{2}+x(1-x)p^{2}
\EE
is a complex, nonreal quantity due to $p_{0}^{2}$ itself being complex with $\tx{Im}(p_{0}^{2})\neq 0$ (here we are assuming that the external momentum $p^{2}\in\Bbb{R}$). The angular integration in Eq.~\eqref{intcc1} can be readily performed, yielding
\begin{align}\label{intcc2}
\nn\mc{I}&=\frac{\Omega_{d-1}}{(2\pi)^{d}}\int_{0}^{1}dx\int_{0}^{+\infty} dq\ q^{d-1}\ \frac{(q^{2})^{n}}{(q^{2}+\Delta)^{2}}=\\
&=\frac{\Omega_{d-1}}{2(2\pi)^{d}}\int_{0}^{1}dx\int_{0}^{+\infty} dy\ \frac{y^{d/2-1+n}}{(y+\Delta)^{2}},
\end{align}
where $\Omega_{d-1}$ is the volume of the $(d-1)$-dimensional unit sphere and on the last line we have changed variable of integration to $y=q^{2}$. The integrand in Eq.~\eqref{intcc2} has a complex pole outside of the domain of integration -- i.e. the positive real axis --, at $y=-\Delta$. The integral over the $y$ variable can be expressed as the limit
\BE
\int_{0}^{+\infty} dy\ \frac{y^{d/2-1+n}}{(y+\Delta)^{2}}=\lim_{\Lambda\to +\infty}\int_{0}^{\Lambda} dy\ \frac{y^{d/2-1+n}}{(y+\Delta)^{2}}.
\EE
We can now change the contour of integration of the definite integral on the right-hand side by setting
\begin{align}\label{intcc3}
&\nn\int_{0}^{\Lambda} dy\ \frac{y^{d/2-1+n}}{(y+\Delta)^{2}}=\oint_{\gamma} dy\ \frac{y^{d/2-1+n}}{(y+\Delta)^{2}}+\\
&\quad-\int_{\gamma_{2}} dy\ \frac{y^{d/2-1+n}}{(y+\Delta)^{2}}-\int_{\gamma_{\Lambda}} dy\ \frac{y^{d/2-1+n}}{(y+\Delta)^{2}},
\end{align}
where $\gamma=\gamma_{1}+\gamma_{\Lambda}+\gamma_{2}$ and the contours $\gamma_{1},\gamma_{\Lambda}$ and $\gamma_{2}$ are displayed in Fig.~20. In particular, $\gamma_{2}$ is chosen so that $y\in\gamma_{2}$ is opposite to $-\Delta$ with respect to the origin of the complex plane. Since the integral over the closed contour $\gamma$ in Eq.~\eqref{intcc3} is zero by analyticity, we have
\BE
\int_{0}^{+\infty} dy\ \frac{y^{d/2-1+n}}{(y+\Delta)^{2}}=\lim_{\Lambda\to +\infty}\int_{-\gamma_{2}} dy\ \frac{y^{d/2-1+n}}{(y+\Delta)^{2}},
\EE
where the integral over $\gamma_{\Lambda}$ drops out in the limit $\Lambda\to +\infty$ \footnote{Keep in mind that, in dimensional regularization, all the integrals are assumed to converge before the limit $d\to 4$ is taken. As a consequence, integrals at infinity such as the one over $\gamma_{\Lambda}$ in Eq.~\eqref{intcc3} can be safely set to zero.}. Moreover, by construction, the argument of $y\in -\gamma_{2}$ satisfies $\tx{arg}(y)=\tx{arg}(\Delta)$. Therefore we can write
\begin{align}
\nn&\int_{0}^{+\infty} dy\ \frac{y^{d/2-1+n}}{(y+\Delta)^{2}}=\\
&=(e^{i\tx{arg}(\Delta)})^{d/2-2+n}\int_{0}^{+\infty} dy\ \frac{y^{d/2-1+n}}{(y+|\Delta|)^{2}}.
\end{align}
One last change of integration variables from $y$ to $y/|\Delta|$ leaves us with
\begin{align}
\nn&\int_{0}^{+\infty} dy\ \frac{y^{d/2-1+n}}{(y+\Delta)^{2}}=\\
\nn&=(|\Delta|e^{i\tx{arg}(\Delta)})^{d/2-2+n}\int_{0}^{+\infty} dy\ \frac{y^{d/2-1+n}}{(y+1)^{2}}=\\
&=\Delta^{d/2-2+n}\ \Gamma(d/2+n)\Gamma(2-d/2-n).
\end{align}
The latter is the very same result found for $\Delta\in\Bbb{R}$. Hence the integral $\mc{I}$ can be computed as if $\Delta$ were a real number or, equivalently, as if $p_{0}^{2}$ were real.

\begin{figure}[t]
\vskip 1cm
\centering
\includegraphics[width=0.40\textwidth]{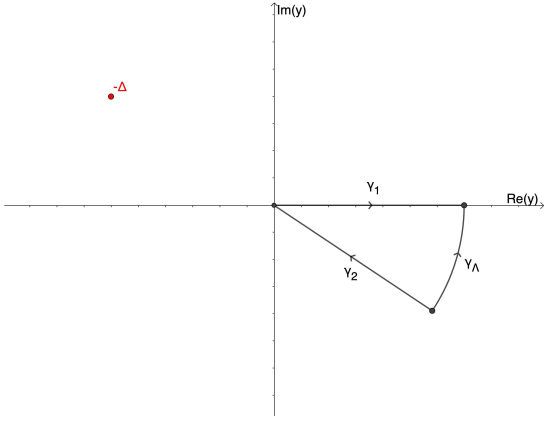}
\caption{Contour for the loop integrals in the CC scheme. $\gamma_{2}$ is chosen so that $y\in \gamma_{2}$ is opposite to the pole $-\Delta$ with respect to the origin of the complex plane, hence $\tx{arg}(y)=\tx{arg}(\Delta)$.}
\end{figure}

Finally, since the diagrams for the CC scheme (Fig.~6) are identical to those of the minimalistic scheme (Fig.~4, diagrams (2a) to (2c)) except for the fact that the internal gluon propagator is made up of two terms, each multiplied by a factor of $R$ or $\overline{R}$, by considering each of these two terms separately we find that
\begin{align}
\nn&\Sigma^{(\tx{loops})}_{\tx{c.c.}}(p)=\\
&=R\,\Sigma^{(\tx{loops})}_{\tx{m.}}(p)\Big|_{m^{2}=p_{0}^{2}}+\overline{R}\,\Sigma^{(\tx{loops})}_{\tx{m.}}(p)\Big|_{m^{2}=\overline{p_{0}^{2}}},
\end{align}
where $\Sigma^{(\tx{loops})}_{\tx{c.c.}}(p)$ and $\Sigma^{(\tx{loops})}_{\tx{m.}}(p)$ are the loop contributions to the 1PI quark self-energies computed, respectively, in the CC scheme and in the minimalistic scheme, and $m^{2}$ is the gluon mass parameter introduced by the screened expansion.

\end{document}